\documentclass[12pt]{article}

\usepackage[square,comma,sort&compress]{natbib}
\usepackage{epsfig}
\usepackage{rotating}

\def\vb#1{\vbox to #1 pt{}}

\DeclareMathAlphabet{\mathsc}{OT1}{cmr}{m}{sc}

\newcommand{\dof}  {d.o.f.}

\newcommand{\eVq}  {\rm{eV}^2}
\newcommand{\Sol}  {\mathsc{sol}}
\newcommand{\Atm}  {\mathsc{atm}}

\def\21{$SU(2) \otimes U(1) $}

\newcommand{\Dms}  {\Delta m^2_\Sol}
\newcommand{\Dma}  {\Delta m^2_\Atm}
\newcommand{\AHEP}{AHEP Group, Instituto de F\'{\i}sica Corpuscular --
  C.S.I.C./Universitat de Val{\`e}ncia \\
  Edificio de Institutos de Paterna, Apartado 22085,
  E--46071 Val{\`e}ncia, Spain\\}

\begin{document}

\begin{titlepage} 

\begin{flushright}
hep-ph/0302021 \\ 
IFIC/03-03\\
ZU-TH02/03 \\ 

\end{flushright} 
\vspace*{3mm} 
\begin{center}  

 \textbf{\large Solar Neutrino Masses and Mixing from Bilinear R-Parity Broken 
Supersymmetry: Analytical versus Numerical Results}\\[10mm]

{M. A. D\'{\i}az${}^1$, M. Hirsch${}^2$,  W. Porod${}^3$, 
J. C. Rom\~ao${}^{2,4}$ and J. W. F. Valle${}^2$ } 
\vspace{0.3cm}\\ 

{\it $^1$ Facultad de F\'\i sica, Universidad Cat\'olica de Chile\\ 
          Av. Vicu\~na Mackenna 4860, Santiago, Chile\\}

{\it $^2$ \AHEP}

{\it $^3$ Institut f\"ur Theoretische Physik, Universit\"at Z\"urich, \\ 
CH-8057 Z\"urich, Switzerland}

{\it $^4$ Departamento de F\'\i sica and CFIF, Instituto Superior T\'ecnico\\
          Av. Rovisco Pais 1, $\:\:$ 1049-001 Lisboa, Portugal \\}

\end{center}

\begin{abstract} 
  We give an analytical calculation of solar neutrino masses and
  mixing at one-loop order within bilinear R-parity breaking
  supersymmetry, and compare our results to the exact numerical
  calculation.  Our method is based on a systematic perturbative
  expansion of R-parity violating vertices to leading order. We find
  in general quite good agreement between approximate and full 
  numerical calculation, but the approximate expressions are much 
  simpler to implement. Our formalism works especially well for 
  the case of the large mixing angle MSW solution (LMA-MSW), now 
  strongly favoured by the recent KamLAND reactor neutrino data. 
\end{abstract} 
 
\end{titlepage}

\newpage

\setcounter{page}{1} 

\section{Introduction}

Solar neutrino experiments, including the measurement of the neutral
current rate for solar neutrinos by the SNO collaboration
\cite{Ahmad:2002jz} provide a solid evidence for solar neutrino
conversions~\cite{Maltoni:2002ni}. This has been recently confirmed by
the first results from the KamLAND experiment using reactor
(anti)-neutrinos~\cite{:2002dm,Pakvasa:2003zv}.
Combining the information from reactors with all of the solar neutrino
data leads to the best fit point~\cite{Maltoni:2002aw}:
\begin{equation}
    \tan^2\theta_\Sol = 0.46, \qquad \Dms = 6.9\times10^{-5}~\eVq \,,
\end{equation}
confirming that the solar neutrino mixing angle is large, but
significantly non-maximal. The $3\sigma$ region for $\theta$ is:
\begin{equation}
\label{thetasol.range}
    0.29 \leq \tan^2\theta_\Sol \leq 0.86,
\end{equation}
based on a combination of all experimental data. However, one finds a
significant reduction of the allowed $\Dms$ range. As shown in
Ref.~\cite{Maltoni:2002aw}, the pre-KamLAND LMA-MSW region is now
split into two sub-regions. At $3\sigma$ (1 dof.) one obtains

\begin{equation}
\label{sol.kam.ranges}
    5.1\times 10^{-5}~\eVq \leq \Dms \leq 9.7\times 10^{-5}~\eVq, \\
    1.2\times 10^{-4}~\eVq \leq \Dms \leq 1.9\times 10^{-4}~\eVq.
\end{equation}

Altogether, KamLAND results exclude all oscillation solutions except
for the large mixing angle MSW solution (LMA-MSW) to the solar
neutrino problem~\cite{Gonzalez-Garcia:1999aj}.

On the other hand, current atmospheric neutrino data require
oscillations involving $\nu_{\mu} \leftrightarrow \nu_{\tau}$
~\cite{Fukuda:1998mi}. The most recent global analysis
gives~\cite{Maltoni:2002ni},
\begin{equation}
\label{t23d23}
    \sin^2\theta_\Atm = 0.5 \,,\: \Dma = 2.5 \times
    10^{-3}~\eVq \: 
\end{equation}
with the 3$\sigma$ ranges (1 \dof)
\begin{eqnarray}
\label{t23d23range} 
    0.3 \le \sin^2\theta_\Atm \le 0.7 \\ 
    1.2 \times 10^{-3}~\eVq \le \Dma  \le 4.8 \times
    10^{-3}~\eVq \,. 
\end{eqnarray}

These data have triggered a rush of theoretical and phenomenological
papers on models of neutrino masses and mixings, most of which
introduce a large mass scale in order to implement various variants of
the see-saw mechanism \cite{seesaw,Mohapatra:1980yp,Schechter:1981cv}.
Broken R-parity supersymmetry provides a theoretically interesting and
phenomenologically viable alternative to the origin of neutrino mass
and mixing~\cite{Ross:1985yg}.  Here we focus on the simplest case of
supersymmetry with bilinear R-parity breaking \cite{Diaz:1997xc}.
In contrast to the seesaw mechanism, here neutrino masses are
generated at the electro-weak scale. Such low-scale schemes for
neutrino masses have the advantage of being testable also
in accelerator experiments~\cite{Bartl:2000yh}-\cite{Hirsch:2002ys} 
through the decay properties of the lightest supersymmetric 
particle if the LSP is a 
neutralino~\cite{Porod:2000hv,Mukhopadhyaya:1998xj,Choi:1999tq}, a 
slepton~\cite{Hirsch:2002ys} 
or a stop~\cite{Restrepo:2001me,Allanach:1999bf}.

Supersymmetric models with explicit bilinear breaking of R-parity
(BRpV)
\cite{deCampos:1995av,Banks:1995by,deCarlos:1996du,Akeroyd:1997iq,
BRpVrecent,BRpVother,BRpV_tau,BRpVmore}
provide a simple and calculable framework for neutrino masses and
mixing angles in agreement with the experimental data
\cite{Hirsch:2000ef}. In this model the atmospheric neutrino mass
scale is generated at tree-level, through an effective `low-scale''
variant of the seesaw mechanism \cite{Ross:1985yg}. In contrast, 
the solar mass and mixings are generated radiatively~\cite{Hirsch:2000ef}.
Tree-level neutrino masses within BRpV have been 
treated extensively in the literature. 

This paper is mainly devoted to the solar neutrino masses and mixing.
An accurate and reliable calculational method is now necessary in
order to confront the model with the new experimental data from
KamLAND and other neutrino experiments. A complete one-loop
calculation of the neutrino-neutralino mass matrix has been
given~\cite{Hirsch:2000ef} but is rather complex. On the other hand,
approximations to the full 1-loop calculation which exist in the 
literature~\cite{Chun:2002vp} have not been tested yet against 
the full calculation. Especially in view of future experimental 
sensitivities we think such a ``benchmark'' is important.

In this paper we give an accurate determination of neutrino mass and
mixing within an analytical approximation and obtain formulae which
can be rather simple, in some cases.  For definiteness we will stick
to the case of explicit BRpV only.  This is the simplest of all R
parity violating models. It can be considered either as a minimal
three--parameter extension of the MSSM (with no new particles) valid
up to some very high unification energy scale, or as the effective
description of a more fundamental theory in which the breaking of
R-parity is spontaneous \cite{Masiero:1990uj,Romao:vu,spo:2}.  The
latter implies the absence of trilinear R-parity breaking parameters
in the superpotential~\footnote{Alternatively, such absence may arise
  from suitable symmetries~\cite{Mira:2000gg}}.

This paper is organized as follows.  In Sec.~\ref{sec:brpv-formalism}
we introduce the main features of the model and the relevant mass
matrices and corresponding diagonalization matrices.  In particular we
identify the relevant Feynman graph topologies and rules, and derive
approximate formulae for the couplings relevant for the determination
of radiatively induced solar neutrino mass scale. We give approximate 
formulas for the bottom quark/squark loop as well as for the charged 
scalar loop. In Sec.~\ref{sec:analyt-vers-numer} we check the accuracy of 
our approximation formulas by a comparison with a full numerical
calculation, studying first the role of the simplest bottom-sbottom loop, 
then the charged scalar loop, before comparing the sum of the two 
to the full numerical result. In \ref{sec:simpl-appr-form} we give 
simplified approximation formulas for the solar mass and solar mixing 
angle and conclude and summarize our results in \ref{sec:disc-concl}.

\section{BRpV Formalism}
\label{sec:brpv-formalism}

In this section we introduce the main features of the model and the
relevant mass matrices, and develop approximate formulas, first for
couplings and then for the radiative contributions to the neutrino
masses due to the exchange of bottom and sbottom quarks, and due to
charged scalars and charged fermion loops.

\subsection{BRpV Model}
\label{sec:brpv-model}

The minimal BRpV model we are working with is characterized by the
presence of three extra bilinear terms in the superpotential analogous
to the $\mu$ term present in the MSSM
\begin{equation}
W=W_{Yuk}+\varepsilon_{ab}\left(-\mu\widehat H_d^a\widehat H_u^b
+\epsilon_i\widehat L_i^a\widehat H_u^b\right)
\end{equation}
where $W_{Yuk}$ includes the usual MSSM Yukawa terms, $\mu$ is the
Higgsino mass term of the MSSM, and $\epsilon_i$ are the three new
terms which violate R-Parity and lepton number. The smallness of
$\epsilon_i$ may arise dynamically (the product of a Yukawa coupling
times a singlet sneutrino vacuum expectation value) in models with
spontaneous breaking of R parity~\cite{Masiero:1990uj}. 

Alternatively, the smallness of the $\epsilon_i$ may arise from
suitable family symmetries~\cite{Mira:2000gg}.  In fact any solution
to the $\mu$ problem~\cite{Giudice:1988yz} potentially explains also
the ``$\epsilon_i$-problem''~\cite{Nilles:1996ij}. In fact a common
origin for the $\epsilon_i$ terms responsible for the explanation of
the neutrino anomalies, and the $\mu$ term accounting for electroweak
symmetry breaking can be ascribed to a suitable horizontal symmetry
that may also predict their ratio, as in~\cite{Mira:2000gg}.

In addition we have the corresponding soft supersymmetry breaking
terms in the scalar potential,
\begin{equation}
V_{soft}=V'_{soft}+\varepsilon_{ab}\left(
-B\mu H_d^a H_u^b+B_i\epsilon_i\widetilde L_i^a H_u^b\right)
\end{equation}
where $B$ and the three $B_i$ have units of mass and in $V'_{soft}$ we
include all the usual mass and trilinear supersymmetry breaking terms
of the MSSM.

\subsection{Rotation Matrices}
\label{sec:rotation-matrices}

If the effective RpV parameters are smaller than the weak scale, we
can work in a perturbative expansion defined by $\xi \ll 1$, where
$\xi$ denotes a $3\times 4$ matrix given as~\cite{Hirsch:1998kc} 

\begin{eqnarray}
\xi_{i1} &=& \frac{g' M_2 \mu}{2\Delta_0}\Lambda_i \cr
\vb{20}
\xi_{i2} &=& -\frac{g M_1 \mu}{2\Delta_0}\Lambda_i \cr
\vb{20}
\xi_{i3} &=& - \frac{\epsilon_i}{\mu} + 
          \frac{M_{\tilde\gamma} v_u}{4\Delta_0}\Lambda_i \cr
\vb{20}
\xi_{i4} &=& - \frac{M_{\tilde\gamma} v_d}{4\Delta_0}\Lambda_i
\label{xielementos}
\end{eqnarray}
where $\Delta_0$ is the determinant of the $4\times 4$ neutralino 
mass matrix, $M_{\tilde\gamma}=g^2M_1+g'^2M_2$ and

\begin{equation}
\Lambda_i = \mu v_i + v_d \epsilon_i 
\label{lambdai}
\end{equation}

The neutralino/neutrino mass matrix is
diagonalized by a $7\times7$ rotation matrix ${\cal N}$ according to
\begin{equation}
{\cal N}^* {\bf M}_{F^0} {\cal N}^{-1}={\bf M}_{F^0}^{\mathrm diag}
\end{equation}
and the eigenvectors are given by
\begin{equation}
F^0_i={\cal N}_{ij}\psi_j
\end{equation}
using the basis $\psi=(-i\lambda',-i\lambda^3,\widetilde{H}_d^1,
\widetilde{H}_u^2, \nu_{e},\nu_{\mu}, \nu_{\tau} )$. In this 
approximation, the rotation matrix can be written as 
\begin{equation}
{\cal N}^*\approx\left(\matrix{
N^* & N^*\xi^{\dagger} \cr
-V_{\nu}^T\xi & V_{\nu}^T 
}\right)
\end{equation}
Here, $N$ is the rotation matrix that diagonalizes the $4\times4$ MSSM
neutralino mass matrix, $V_{\nu}$ is the rotation matrix that
diagonalizes the tree level neutrino $3\times3$ mass matrix, and
$\xi_{ij}\ll 1$ are the expansion parameters \cite{Hirsch:1998kc,Now96}.
The terms we need are
\begin{equation}
V_{\nu}^T\xi=\left(\matrix{
0 & 0 & b\tilde\epsilon_1 & 0 \cr
0 & 0 & b\tilde\epsilon_2 & 0 \cr
a_1|\vec\Lambda| & a_2|\vec\Lambda| & a_3|\vec\Lambda|+b\tilde\epsilon_3 
& a_4|\vec\Lambda|
}\right)
\label{Rnuxi}
\end{equation}
where $b=-1/\mu$,
\begin{eqnarray}
a_1={{g'M_2\mu}\over{2\Delta_0}}\,,\quad
a_2=-{{gM_1\mu}\over{2\Delta_0}}\,,\quad
a_3={{M_{\tilde\gamma}v_u}\over{4\Delta_0}}\,,\quad
a_4=-{{M_{\tilde\gamma}v_d}\over{4\Delta_0}}\,,
\end{eqnarray}
The $\tilde\epsilon$ parameters in eq.~(\ref{Rnuxi}) are defined as
$\tilde\epsilon_i=\left(V_{\nu}^T\right)^{ij}\epsilon_j$, and are
given by
\begin{eqnarray}
\label{eq:TildeEpsilon}
\tilde\epsilon_1&=&{{
\epsilon_e(\Lambda_{\mu}^2+\Lambda_{\tau}^2)-\Lambda_e
(\Lambda_{\mu}\epsilon_{\mu}+\Lambda_{\tau}\epsilon_{\tau})
}\over{
\sqrt{\Lambda_{\mu}^2+\Lambda_{\tau}^2}
\sqrt{\Lambda_e^2+\Lambda_{\mu}^2+\Lambda_{\tau}^2}
}}
\nonumber\\
\tilde\epsilon_2&=&{{
\Lambda_{\tau}\epsilon_{\mu}-\Lambda_{\mu}\epsilon_{\tau}
}\over{
\sqrt{\Lambda_{\mu}^2+\Lambda_{\tau}^2}
}}
\\
\tilde\epsilon_3&=&{{
\vec\Lambda\cdot\vec\epsilon
}\over{
\sqrt{\Lambda_e^2+\Lambda_{\mu}^2+\Lambda_{\tau}^2}
}}\nonumber
\end{eqnarray}

On the other hand the chargino/charged slepton mass matrix is
diagonalized with two different $5\times5$ mass matrices,
\begin{equation}
{\cal U}^*\, {\bf M}_{F^+} {\cal V}^{-1}={\bf M}_{F^+}^{\mathrm diag}
\end{equation}
with the eigenvectors satisfying
\begin{equation}
F^+_{Ri}={\cal V}_{ij}\psi^+_j\,,\qquad 
F^-_{Li}={\cal U}_{ij}\psi^-_j
\end{equation}
in the basis 
$\psi^+=(-i\lambda^+,\widetilde H_2^1,e_R^+,\mu_R^+,\tau_R^+)$ and
$\psi^-=(-i\lambda^-,\widetilde H_1^2,e_L^-,\mu_L^-,\tau_L^-)$, and with
the Dirac fermions being
\begin{equation}
F_i^+=\left(\matrix{F^+_{Ri} \cr \cr \overline{F^-_{Li}}}\right)
\end{equation}
To first order in the R-Parity violating parameters we have
\begin{equation}
{\cal V}\approx\left(\matrix{
V & V\xi^{T}_R \cr
-V^{\ell}_R\xi^{*}_R & V^{\ell}_R
}\right)\,,\qquad
{\cal U}\approx\left(\matrix{
U & U\xi^{\dagger}_L \cr
-V^{{\ell}*}_L\xi_L & V^{{\ell}*}_L
}\right)
\end{equation}
where $V^{{\ell}*}_L$ and $V^{\ell}_R$ diagonalize the charged lepton
mass matrix according to $V^{{\ell}*}_L{\bf
  M}^{\ell}V^{\ell\dagger}_R={\bf M}^{\ell}_{\mathrm diag}$.  For the
purposes of our approximate formula, it is sufficient to take 
$\xi_R={\bf 0}_{2\times3}$, because the mixing between right-handed 
leptons and the charginos is supressed with respect to $\xi_L$ by 
a factor of $m_l/M_{SUSY}$ \cite{Hirsch:1998kc,Now96}.
Note, that we can choose $V^{{\ell}*}_L = 
V^{\ell\dagger}_R = {\bf 1}_{3\times 3}$. We then have
\begin{equation}
\xi_L^{i1}=a^L_1\Lambda_i\,,\qquad \xi_L^{i2}=a^L_2\Lambda_i+b\epsilon_i
\end{equation}
and
\begin{equation}
a_1^L={g\over{\sqrt{2}\Delta_+}}\,,\qquad 
a_2^L=-{{g^2v_u}\over{2\mu\Delta_+}}
\end{equation}
where $\Delta_+$ is the determinant of the $2\times2$ chargino mass matrix.

In the BRpV model the charged Higgs fields mix with the charged
sleptons forming an $8\times8$ mass matrix \cite{Hirsch:2000ef}, which
is diagonalized by a rotation matrix ${\bf R}_{S^{\pm}}$. The 
construction of ${\bf R}_{S^{\pm}}$ to first order in small (RpV) 
parameters is quite straightforward but lengthy. The interested 
reader can find the details in Appendix~\ref{ap:Rotations}.

\subsection{Approximate Couplings}
\label{sec:appr-coupl}

The relevant Feynman rules for the bottom-sbottom loops are, in the
case of left sbottoms:
\begin{center}
\includegraphics{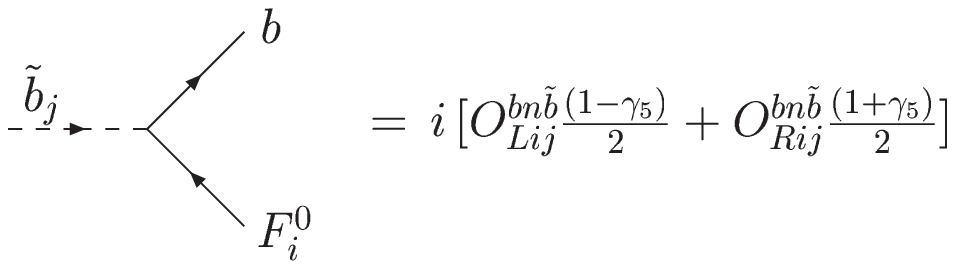}
\end{center}
with
\begin{eqnarray}
O^{bn\tilde b}_{Lij}&=&-R^{\tilde b}_{j1}h_b{\cal N}^*_{i3}
-R^{\tilde b}_{j2}{{2g}\over{3\sqrt{2}}}t_W{\cal N}^*_{i1}
\nonumber\\
O^{bn\tilde b}_{Rij}&=&R^{\tilde b}_{j1}{g\over{\sqrt{2}}}
\left({\cal N}_{i2}-{\textstyle{1\over3}}t_W{\cal N}_{i1}\right)
-R^{\tilde b}_{j2}h_b{\cal N}^*_{i3}
\label{Obnsb}
\end{eqnarray}
where $t_W=\tan\theta_W$. After approximating the rotation matrix 
${\cal N}$ we find that expressions similar to eq.~(\ref{Obnsb}) with
the replacement ${\cal N}\rightarrow{\mathrm N}$ are valid when the
neutral fermion is a neutralino. When the neutral fermion $F^0$ is a 
neutrino, the following expressions hold
\begin{eqnarray}
O^{bn\tilde b}_{Lij} &\approx& R^{\tilde b}_{j1}h_b\left(
a_3|\vec\Lambda|\delta_{i'3}+b\tilde\epsilon_{i'}\right)
+R^{\tilde b}_{j2}{{2g}\over{3\sqrt{2}}}t_Wa_1|\vec\Lambda|\delta_{i'3}
\nonumber\\
O^{bn\tilde b}_{Rij} &\approx& R^{\tilde b}_{j1}{g\over{\sqrt{2}}}
\left({\textstyle{1\over3}}t_Wa_1-a_2\right)|\vec\Lambda|\delta_{i'3}
+R^{\tilde b}_{j2}h_b\left(a_3|\vec\Lambda|\delta_{i'3}+
b\tilde\epsilon_{i'}\right)
\end{eqnarray}
where $i'=i-4$ label one of the neutrinos. $R^{\tilde b}_{jk}$ are the 
rotation matrices connecting weak and mass eigenstate basis for the 
scalar bottom quarks. In case of no intergenerational mixing in the 
squark sector $R^{\tilde b}_{jk}$ can be parameterized by just one 
diagonalizing angle $\theta_{\tilde b}$.

The relevant Feynman rule for the charged Higgs/slepton loops is
\begin{center}
\includegraphics{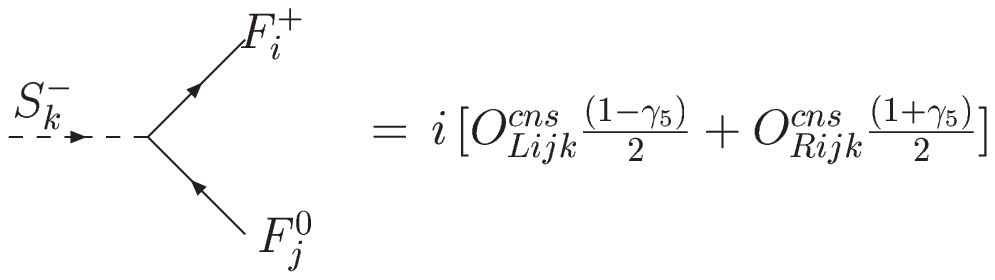}
\end{center}
where the $O^{cns}_{Lijk}$ and $O^{cns}_{Rijk}$ couplings are given in
Appendix~\ref{ap:couplings} in Eqs.~(\ref{eq:ocnsL}, \ref{eq:ocnsR}).

After approximating the rotation matrices ${\cal U}$, and ${\cal V}$
in the chargino sector, and ${\cal N}$ in the neutralino sector we
find approximate expressions for these couplings that we will use
below. These formulae are collected in Eqs.~(\ref{eq:charginoL},
\ref{eq:leptonL}, \ref{eq:charginoR}, \ref{eq:leptonR}) of
Appendix~\ref{ap:couplings}.

\subsection{Relevant Topologies}
\label{sec:relevant-topologies}

We now give the structure of the mass matrices relevant for the
determination of solar neutrino masses and mixings. While in the BRpV
model the atmospheric anomaly is explained at the tree-level, the
solar neutrino masses and mixings are both generated
radiatively. In particular, the ``solar angle'' has no meaning at the
tree level due to the degeneracy of the two lightest neutrinos in this
limit.

Diagonalizing the tree-level neutrino mass matrix first and adding 
then the 1-loop corrections before re-diagonalization the resulting 
neutrino/neutralino mass matrix has non-zero entries in the 
neutrino/neutrino, the neutrino/neutralino and in the neutralino/neutralino 
sectors. We have found that the most important part of the 1-loop 
neutrino masses derives from the neutrino/neutrino sector and that 
the 1-loop induced neutrino/neutralino mixing is usually subdominant.

\begin{figure}[ht]
  \begin{center}
  \begin{tabular}{ccc}
    \includegraphics[width=0.28\linewidth]{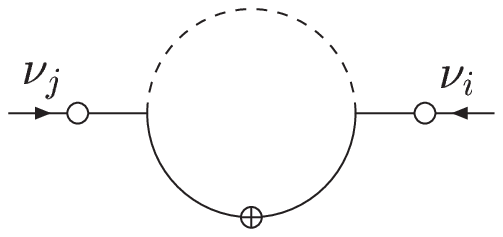}&
    \includegraphics[width=0.28\linewidth]{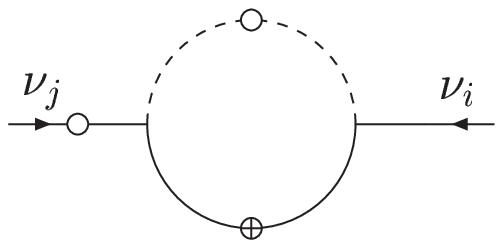}&
    \includegraphics[width=0.28\linewidth]{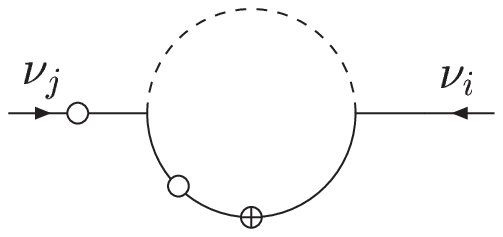}
  \end{tabular}
  \begin{tabular}{ccc}
    \includegraphics[width=0.28\linewidth]{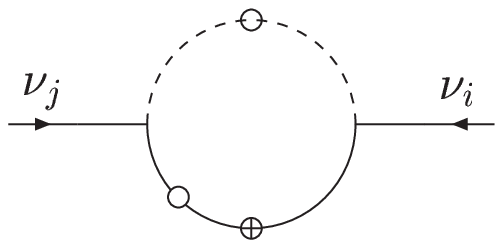}&
     & \includegraphics[bb= 115 628 260 678,height=15mm]{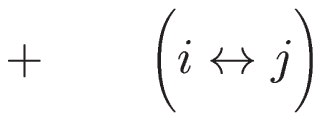}
  \end{tabular}
\caption{Topologies for neutrino self-energies in the BRpV 
supersymmetric model}
\label{fig:topologies}
\end{center}
\end{figure}

The relevant topologies for the one loop calculation of neutrino
masses are then illustrated in Fig.\ref{fig:topologies}.
Here our conventions are as follows: open circles with a cross inside
indicate genuine mass insertions which flip chirality.  On the other
hand open circles without a cross correspond to small R-Parity
violating projections, indicating how much of an Rp-even/odd mass
eigenstate is present in a given Rp-odd/even weak eigenstate.
Strictly speaking these projections are really coupling matrices
attached to the vertices, and this is what appears in the numerical
code.  However, given the smallness of Rp-violating effects, the
``insertion-method'' proves to be a rather useful tool to develop an
analytical perturbative expansion and to acquire some simple
understanding of the results.

\subsection{Bottom-sbottom loops}
\label{sec:bottom-sbottom-loops}

\begin{figure}[ht]
\begin{center}
\begin{tabular}{cc}
\includegraphics[width=0.45\linewidth]{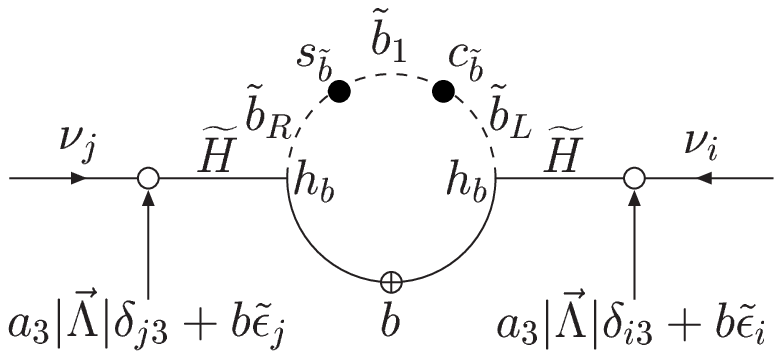}
&\includegraphics[width=0.45\linewidth]{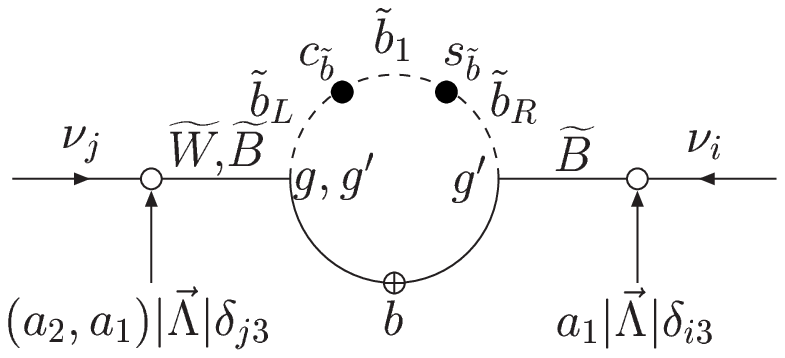}
\end{tabular}
\end{center}
\vspace{-3mm}
\caption{Bottom--Sbottom diagrams for solar neutrino mass in the BRpV model}
\label{fig:BottomSbottomLoop}
\end{figure}

The simplest contribution to the radiatively induced neutrino mass
arises from loops involving bottom quarks and squarks is given by
\cite{Hirsch:2000ef}
\begin{equation}
\widetilde\Pi_{ij}(0)=-{{N_c}\over{16\pi^2}}\sum_r\left(
O^{bn\tilde b}_{Rjr}O^{bn\tilde b}_{Lir}+
O^{bn\tilde b}_{Ljr}O^{bn\tilde b}_{Rir}\right)
m_bB_0(0,m_b^2,m_r^2)
\end{equation}
$B_0(0,m_b^2,m_r^2)$ is the usual Passarino-Veltman 
function \cite{Passarino:1978jh,FFLP}. 
This contribution can be expressed as being proportional to the
difference of two $B_0$ functions, 
\begin{equation}
  \label{eq:PV}
  \Delta B_0^{\tilde b_1\tilde b_2}=
  B_0(0,m_b^2,m_{\tilde b_1}^2)-B_0(0,m_b^2,m_{\tilde b_2}^2)
\end{equation}
as follows
\begin{equation}
\label{eq:bsb}
\Delta\widetilde\Pi_{ij}=-{{N_c m_b}\over{16\pi^2}}
2s_{\tilde b}c_{\tilde b}h_b^2
\Delta B_0^{\tilde b_1\tilde b_2}
\left[
 {{\tilde\epsilon_i\tilde\epsilon_j}\over{\mu^2}}
  + a_3 b \left(\tilde\epsilon_i\delta_{j3}+
 \tilde\epsilon_j\delta_{i3}\right)|\vec\Lambda|
 + \left( a_3^2 + \frac{ a_L a_R}{h^2_b}\right)
 \delta_{i3}\delta_{j3} 
 |\vec\Lambda|^2 \right] 
 \end{equation}
where we have defined
\begin{equation}
a_R={g\over{\sqrt{2}}}\left({1\over3}t_Wa_1-a_2\right)
\,,\qquad
a_L={g\over{\sqrt{2}}}\,{2\over3}t_Wa_1
\end{equation}
The different contributions can be understood as coming from the
graphs corresponding to the first topology of
Fig.~\ref{fig:topologies}. They have been depicted in more detail in
Fig.~\ref{fig:BottomSbottomLoop},
where we have adopted the following conventions: a) as before, open
circles correspond to small R-parity violating projections, indicating
how much of a weak eigenstate is present in a given mass eigenstate,
(b) full circles correspond to R-parity conserving projections and (c)
open circles with a cross inside indicate genuine mass insertions
which flip chirality.

The open and full circles should really appear at the vertices since
the particles propagating in the loop are the mass eigenstates. We
have however separated them to better identify the origin of the
various terms.
There is another set of graphs analogous to the previous ones which
corresponds to the heavy sbottom. They are obtained from the previous
graphs making the replacement $\tilde b_1\rightarrow\tilde b_2$,
$s_{\tilde b}\rightarrow c_{\tilde b}$ and $c_{\tilde b}\rightarrow
-s_{\tilde b}$.  Note that for all contributions to the $2\times2$
submatrix corresponding to the light neutrinos the divergence from
$B_0(0,m_b^2,m_{\tilde b_1}^2)$ is canceled by the divergence from
$B_0(0,m_b^2,m_{\tilde b_2}^2)$, making finite the contribution from
bottom-sbottom loops to this submatrix, as it should be, since the
mass is fully ``calculable''.

\subsection{Charged Scalar-Charged Fermion Loops}
\label{sec:ChargedScalarLoops}

\begin{figure}  \centering
\begin{tabular}{cc} 
\includegraphics[width=0.45\linewidth]{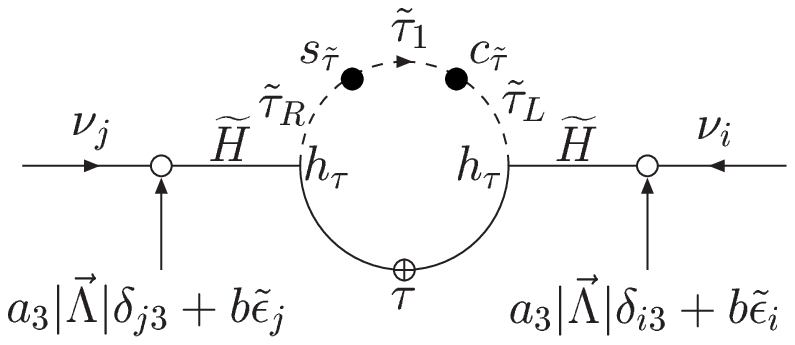}
&\includegraphics[width=0.45\linewidth]{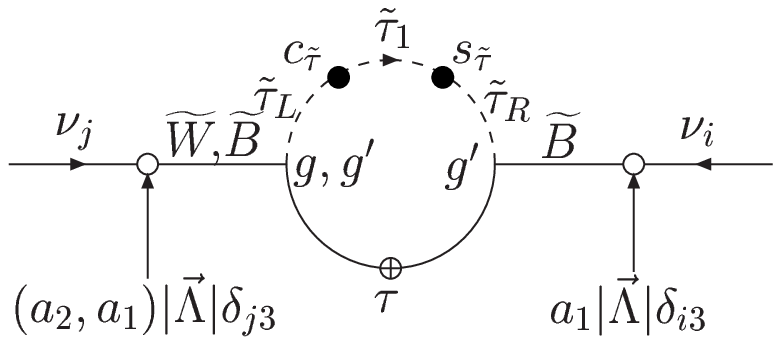}
\end{tabular}
\begin{tabular}{cc}  \centering
\includegraphics[width=0.45\linewidth]{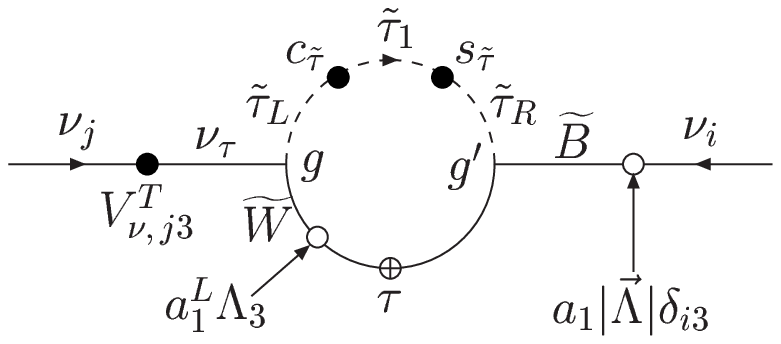}
&\includegraphics[width=0.45\linewidth]{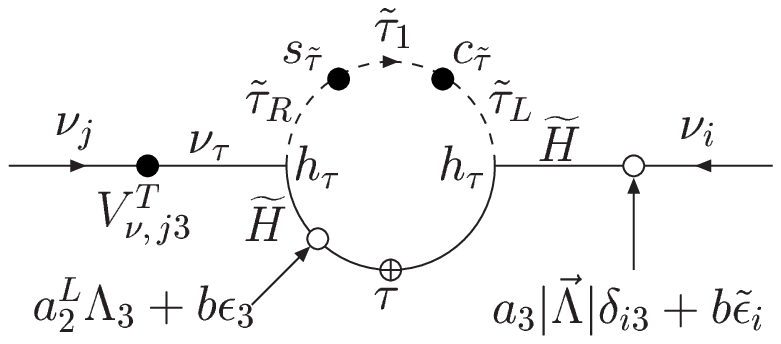}
\end{tabular}
\vspace{-3mm}
\caption{Charged scalar contributions to solar neutrino masses in BRpV 
  model: terms proportional to $\Delta B_0^{\tilde\tau_2\tilde\tau_1}$.}
\label{fig:ChargedScalarLoop}
\end{figure}

Another contribution to the radiatively induced neutrino mass comes
from charged-scalar/charged-fermion loops, given as \cite{Hirsch:2000ef}
\begin{equation}
\widetilde\Pi_{ij}(0)=-{1\over{16\pi^2}}\sum_{k,r}\left(
O^{cns}_{Rkjr}O^{cns}_{Lkir}+O^{cns}_{Lkjr}O^{cns}_{Rkir}\right)
m_kB_0(0,m_k^2,m_r^2)
\end{equation}
The structure of the contribution from charged Higgs/slepton loops is
substantially more complex than that of the bottom-sbottom loop
considered above. It can be expressed as
\begin{eqnarray}
\hskip -3mm
\Delta\widetilde\Pi_{ij}
&\hskip -2mm=\hskip -2mm&
{{m_{\tau}}\over{16\pi^2}}
\left[ \vb{18}
C_{ij}^{\tilde\tau_2\tilde\tau_1} \Delta B_0^{\tilde\tau_2\tilde\tau_1}
+C_{ij}^{H^{\pm}\tilde\tau_1} \Delta B_0^{H^{\pm}\tilde\tau_1}
+C_{ij}^{H^{\pm}\tilde\tau_2} \Delta B_0^{H^{\pm}\tilde\tau_2}
\right. \nonumber\\[+2pt]
&&\hskip 12mm
+C_{ij}^{H^{\pm}{\tilde L}_1} \Delta B_0^{H^{\pm}L_1}
+C_{ij}^{H^{\pm}{\tilde L}_2} \Delta B_0^{H^{\pm}L_2}
+C_{ij}^{G^{\pm}{\tilde L}_1} \Delta B_0^{G^{\pm}L_1}\nonumber\\[+2pt]
&&\hskip 12mm
+C_{ij}^{G^{\pm}{\tilde L}_2} \Delta B_0^{G^{\pm}L_2}
+C_{ij}^{G^{\pm}\tilde\tau_1\tilde\tau_2} \Delta
B_0^{G^{\pm}\tilde\tau_1\tilde\tau_2}
+C_{ij}^{G^{\pm}H^{\pm}\tilde\tau_1\tilde\tau_2} \Delta
B_0^{G^{\pm}H^{\pm}\tilde\tau_1\tilde\tau_2}\nonumber\\[+2pt]
&&\hskip 12mm\left. \vb{14}
+(i\leftrightarrow j)
\right]
  \label{ChargedHiggsLoop}
\end{eqnarray}
where
\begin{eqnarray}
  \label{eq:DB0}
  &&\Delta B_0^{X Y}\equiv
  B_0(0,m_{\tau}^2,m^2_X)-B_0(0,m_{\tau}^2,m_Y^2)
\ ; \
X,Y=(G^{\pm}, H^{\pm}, L_1, L_2, \tilde\tau_1, \tilde\tau_2)
\nonumber\\[+2pt]
&&\Delta B_0^{G^{\pm}\tilde\tau_1\tilde\tau_2}\equiv
c_{\tilde\tau}^2\,
  B_0(0,m^2_{\tau},m^2_{\tilde\tau_1})+s_{\tilde\tau}^2\, 
B_0(0,m^2_{\tau},m^2_{\tilde\tau_2})
-B_0(0,m^2_{\tau},m^2_{G^{\pm}})\nonumber\\[+2pt]
&&
\Delta B_0^{G^{\pm}H^{\pm}\tilde\tau_1\tilde\tau_2}\equiv
c^2_{\beta}\, B_0(0,m^2_{\tau},m^2_{G^{\pm}})
+s^2_{\beta\,} B_0(0,m^2_{\tau},m^2_{H^{\pm}})\\[+2pt]
&&\hskip 26mm
-c_{\tilde\tau}^2\, B_0(0,m^2_{\tau},m^2_{\tilde\tau_1})
-s_{\tilde\tau}^2\, B_0(0,m^2_{\tau},m^2_{\tilde\tau_2})\nonumber
\end{eqnarray}
and
\begin{eqnarray}
  \label{eq:ChargedScalarsCoefficents}
&&
C_{ij}^{\tilde\tau_2\tilde\tau_1}=s_{\tilde\tau}c_{\tilde\tau} 
\left\{\vb{12}
\sqrt{2} g'a_1|\vec\Lambda|\left[gV^T_{\nu,j3}a_1^L\Lambda_3
 -{\textstyle{1\over{\sqrt{2}}}}(ga_2+g'a_1)|\vec\Lambda|\delta_{j3}
 \right]\delta_{i3}\phantom{\hskip 30mm}\right. \nonumber\\[+2pt]
&&\left.\hskip 22mm\vb{12}
 +h_{\tau}^2\left(b\tilde\epsilon_i+a_3|\vec\Lambda|\delta_{i3}
 -c_{\beta}{\textstyle{{v_3}\over{v}}}V^T_{\nu,i3}
 \right)\left[b\tilde\epsilon_j+a_3|\vec\Lambda|\delta_{j3}
 -V^T_{\nu,j3}(a_2^L\Lambda_3+b\epsilon_3)\right]\right\}
\nonumber\\[+2pt]
&&
C_{ij}^{H^{\pm}\tilde\tau_1}=
 -s_{\beta}\Theta_{HL_3}\left\{\vb{12}
 c_{\tilde\tau}h_{\tau}V^T_{\nu,i3}\Big[gV^T_{\nu,j3}a_1^L\Lambda_3
 -{\textstyle{1\over{\sqrt{2}}}}(ga_2+g'a_1)|\vec\Lambda|\delta_{j3}\Big]
\right.
 \nonumber\\[+2pt]
&&\left. \hskip 35mm \vb{12}
 +s_{\tilde\tau}h_{\tau}^2V^T_{\nu,i3}
 \Big[b\tilde\epsilon_j+a_3|\vec\Lambda|\delta_{j3}
 -V^T_{\nu,j3}(a_2^L\Lambda_3+b\epsilon_3)\Big]\right\}
\nonumber\\[+2pt]
&&
C_{ij}^{H^{\pm}\tilde\tau_2}=
 s_{\beta}\Theta_{HR_3}\left\{\vb{12}
 s_{\tilde\tau}h_{\tau}V^T_{\nu,i3}\Big[gV^T_{\nu,j3}a_1^L\Lambda_3
 -{\textstyle{1\over{\sqrt{2}}}}(ga_2+g'a_1)|\vec\Lambda|\delta_{j3}\Big]
\right.
 \nonumber\\[+2pt]
&&\left. \hskip 30mm \vb{12}
 -c_{\tilde\tau}h_{\tau}^2V^T_{\nu,i3}
 \Big[b\tilde\epsilon_j+a_3|\vec\Lambda|\delta_{j3}
 -V^T_{\nu,j3}(a_2^L\Lambda_3+b\epsilon_3)\Big]\right\}
\nonumber\\[+2pt]
&&
C_{ij}^{H^{\pm}L_1}=
-s_{\beta}\widetilde \Theta_{HL_1}\,
 h_{\tau} g V^T_{\nu,i3} V^T_{\nu,j1}a_1^L\Lambda_3
\nonumber\\[+2pt]
&&
C_{ij}^{H^{\pm}L_2}=
-s_{\beta}\widetilde \Theta_{HL_2}\,
 h_{\tau} g V^T_{\nu,i3} V^T_{\nu,j2}a_1^L\Lambda_3
\nonumber\\[+2pt]
&&
C_{ij}^{G^{\pm}L_1}=
-c_{\beta}{{{v_1}\over v}}\,
h_{\tau}gV^T_{\nu,i3}V^T_{\nu,j1}a_1^L\Lambda_3
\\[+2pt]
&&
C_{ij}^{G^{\pm}L_2}=
-c_{\beta}{{{v_2}\over v}}\,
h_{\tau}gV^T_{\nu,i3}V^T_{\nu,j2}a_1^L\Lambda_3
\nonumber\\[+2pt]
&&
C_{ij}^{G^{\pm}\tilde\tau_1\tilde\tau_2}=
c_{\beta}{{{v_3}\over v}}\,
 h_{\tau}V^T_{\nu,i3}\left[\vb{12}
gV^T_{\nu,j3}a_1^L\Lambda_3
 -{{1\over{\sqrt{2}}}}(ga_2+g'a_1)|\vec\Lambda|\delta_{j3}
\right]
\nonumber\\[+2pt]
&&
C_{ij}^{G^{\pm}H^{\pm}\tilde\tau_1\tilde\tau_2}=
h_{\tau} g \tilde\epsilon_i V^T_{\nu,i3} b a_1^L\Lambda_3
\nonumber
\end{eqnarray}
The result of Eq.~(\ref{ChargedHiggsLoop}) can be represented
graphically for better understanding.  The terms proportional to
$\Delta B_0^{\tilde\tau_2\tilde\tau_1}$ come from the graphs of
Fig.~\ref{fig:ChargedScalarLoop}. There is another set of four graphs
corresponding to $\tilde\tau_2$. These are found after making the
replacements $\tilde\tau_1\rightarrow\tilde\tau_2$,
$s_{\tilde\tau}\rightarrow c_{\tilde\tau}$, and
$c_{\tilde\tau}\rightarrow -s_{\tilde\tau}$.
The diagrams in the first row are the ones that are equivalent to
those in the bottom-sbottom loop. They have as characteristic feature
the presence of two Rp-violating insertions (open circles) in the
external legs. However, in contrast to the quark sector, R-parity
violation can also appear in the charged internal lines running in the
loops, since it occurs in the charged fermion sector.  This explains
the origin of the second row in Fig.~\ref{fig:ChargedScalarLoop}.  The
presence of R-parity violating insertions in the internal lines of the
second row in Fig.~\ref{fig:ChargedScalarLoop} correspond to the
second topology in Fig.~\ref{fig:topologies}. The full diagrammatic
explanation of the rest of the terms appearing in
Eq.~(\ref{ChargedHiggsLoop}) is given in detail in
Appendix~~\ref{ap:diagrams}.

\section{Analytical versus Numerical results}
\label{sec:analyt-vers-numer}

In this section we will check the accuracy of the approximation
formulas given in Sec.~\ref{sec:bottom-sbottom-loops} and
\ref{sec:ChargedScalarLoops}. We do this by comparing the results
obtained with their use with a full numerical calculation of the
one-loop contributions to the neutrino mass, whose details can be
found in Ref.~\cite{Hirsch:2000ef}.

As will be explained in more detail below, the relative importance of
the various loops depends on the - currently unknown - supersymmetric
parameters. In order to reduce the number of free parameters in the
following we will adopt the minimal constrained supergravity (mSUGRA)
version of the MSSM.  As a rule of thumb it can be said that the
bottom-sbottom loop usually gives the main contribution to the
neutrino mass matrix when the neutralino is the LSP. On the other
hand, if the scalar tau is the LSP, both bottom-sbottom and charged
scalar loops are of approximately comparable magnitudes.

We have therefore constructed two different random scans over SUSY
parameter space. Both sets start with the following rather generous
parameter ranges: $M_2$ from [0,1.2] TeV, $|\mu|$ from [0,2.5] TeV,
$m_0$ in the range [0,1.0] TeV, $A_0/m_0$ and $B_0/m_0$ [-3,3] and
$\tan\beta$ [2.5,10]. All randomly generated points were subsequently
tested for consistency with the minimization (tadpole) conditions of
the Higgs potential, as well as for phenomenological constraints from
supersymmetric particle searches. We then selected points in which a)
the lightest neutralino is the LSP (called set ``Ntrl'' in the
following) or b) at least one of the charged sleptons was the LSP
(called set ``Stau'' in the following). Note that in the Stau set $m_0
<<M_2$ and large $\mu$ values are strongly preferred.

R-parity violating parameters are chosen in such a way that neutrino
oscillation data are reproduced approximately. As discussed in the
introduction, atmospheric neutrino experiments require a
near-to-maximal atmospheric mixing angle $\theta_\Atm$, with $\Dma$ in
the range given in Eq.~(\ref{t23d23range}). On the other hand reactor
data constrain the electron-neutrino component in the third mass
eigenstate to be small.  And, finally, in combination with solar
neutrino data, the KamLAND data require a $\theta_\Sol$ in the range
given in Eq.~(\ref{thetasol.range}) with $\Dms$ as given in
Eq.~(\ref{sol.kam.ranges}). The latter ranges belong to the LMA-MSW
region indicated by a solar-only global analysis of neutrino data
given in Ref.~\cite{Maltoni:2002ni}.  For completeness we also include
the (pre-KamLAND) LOW and VAC-type solutions of the solar neutrino
anomaly.
In the following we will first discuss the bottom-sbottom and the
charged scalar loops separately, before considering a calculation
taking into account both loops in comparison to the full calculation.

\subsection{Bottom-sbottom loop}
\label{sec:bottom-sbottom-loop}

\begin{figure}[htbp]
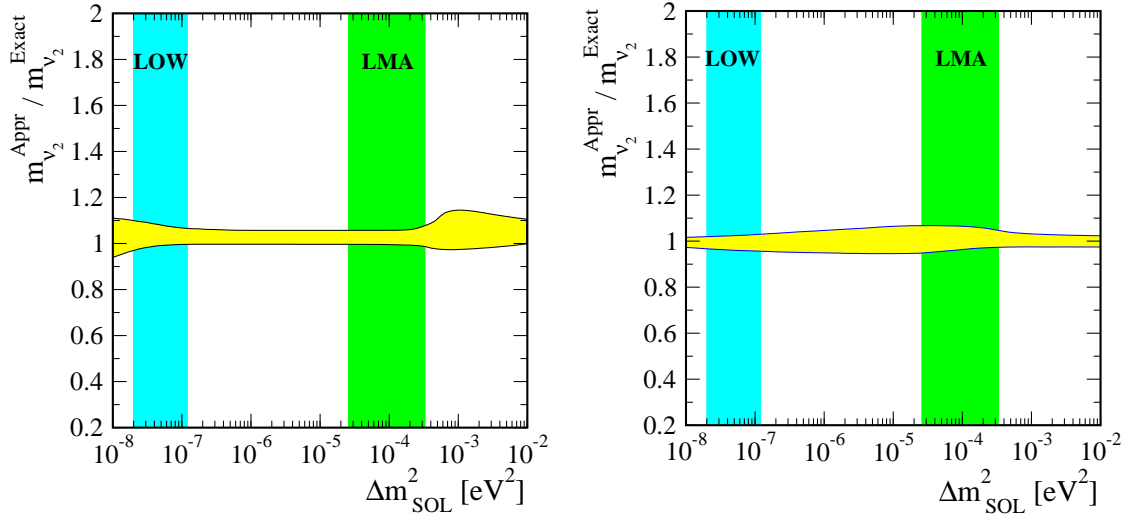
  \centering
  \begin{tabular}{cc}
    \includegraphics[width=0.45\linewidth]{Nfnd1-v4.eps}
    &\includegraphics[width=0.45\linewidth]{Stau4and1-v4.eps}
  \end{tabular}
\caption[]{Ratio $m_{\nu_2}^{Appr}/m_{\nu_2}^{exact}$ versus 
  $\Dms$ in $eV^2$ for the sets Ntrl (left) and Stau (right), for a
  calculation involving only the bottom-sbottom loop. The
  vertical grid lines indicate the 90 \% c.l. regions for the LOW
  and LMA solutions to the solar neutrino problem. }
\label{fig:nd1}
\end{figure}

In Fig.~\ref{fig:nd1} we show the ratio of the approximate-over-exact
solar neutrino mass parameter $m_{\nu_2}^{Appr}/m_{\nu_2}^{exact}$
versus $\Dms$ for the case in which only the bottom-sbottom loop is
taken into account, both in the approximate and in the exact
calculation. The horizontal bands indicate attainable neutrino mass
values when the parameters are scanned as indicated previously. As can
be seen from the figure the approximate formula works quite well for
points in both Ntrl and Stau sets, as long as the neutrino masses fall
in the LMA-MSW range indicated by the right vertical bands.  Note that
the LMA-MSW and LOW bands indicated in the figure correspond to the
full analysis of solar data only, presented in
Ref.~\cite{Maltoni:2002ni}.  The recent KamLAND reactor neutrino data
rule out the LOW solution and restricts the LMA-MSW to somewhat
narrower ranges indicated in Eq.~(\ref{sol.kam.ranges}). One finds
that the mass values inferred from our present analytical
approximation are always within 10 \% or less of the exact numerical
calculation of the bottom-sbottom loop.  Larger deviations show up
only in the Ntrl set, for very small neutrino masses, which we trace
to the neglection of the 1-loop neutrino/neutralino mixing terms in
our approximate treatment. Although not strictly ruled out by a
solar-only global neutrino data analysis~\cite{Maltoni:2002ni}, these
LOW and VAC-type solutions are now strongly disfavored by the latest
KamLAND reactor neutrino data.

\subsection{Charged scalar loop}
\label{sec:charged-scalar-loop}

\begin{figure}
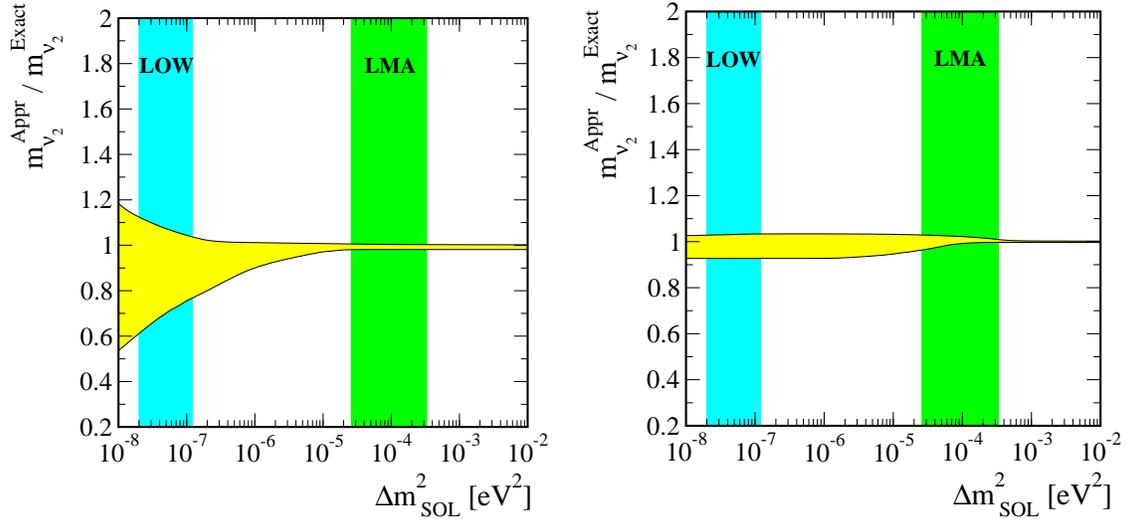
 \centering
  \begin{tabular}{cc}
    \includegraphics[width=0.45\linewidth]{Nfnd3-v4.eps}
    &\includegraphics[width=0.45\linewidth]{Stau4and3-v4.eps}
  \end{tabular}
\caption[]{ ($m_{\nu_2}^{Appr}/m_{\nu_2}^{exact}$) versus 
$\Dms$ [$eV^2$] for the sets Ntrl (left) and Stau 
(right), for a calculation involving only the charged scalar 
loop. }
\label{fig:nd3}
\end{figure}

In Fig.~\ref{fig:nd3} we show the ratio of the approximate-over-exact
solar neutrino mass parameter $m_{\nu_2}^{Appr}/m_{\nu_2}^{exact}$
plotted versus $\Dms$, for a calculation which takes into
account only the charged scalar loop in both the approximate and the
exact calculation.  As can be seen from the figure the approximate
formula is accurate for all points in the LMA-MSW region, indicated by
the right vertical bands~\cite{Maltoni:2002ni}, both for the Stau and
for the Ntrl sets. The only case where our analytic results gives 
a poorer approximation (to better than a factor-of-2) of the full
numerical result is for the Ntrl set, when the neutrino mass falls in
the LOW or VAC ranges, now strongly disfavored by the KamLAND results. 
We have checked numerically that for these very small neutrino masses 
all terms in eq. (\ref{ChargedHiggsLoop}) are of approximately equal 
importance and there are significant cancellations among terms, 
which leads to a less reliable final result.

\subsection{Comparison with full calculation}
\label{sec:comparison-with-full}

\begin{figure}
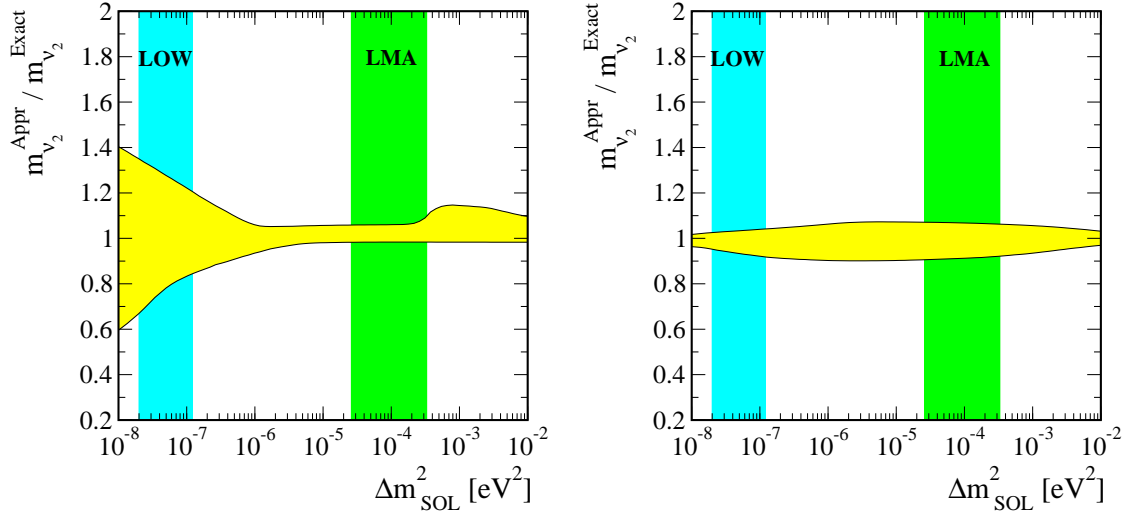
\centering
  \begin{tabular}{cc}
  \includegraphics[width=0.45\linewidth]{NfDelM-v4.eps}
  &\includegraphics[width=0.45\linewidth]{Stau4aDelM-v4.eps}
  \end{tabular}
\caption[] {($m_{\nu_2}^{Appr}/m_{\nu_2}^{exact}$) versus 
  $\Dms$ [$eV^2$] for the set Ntrl (left) and the set Stau (right). 
  $m_{\nu_2}^{Appr}$ is the sum of
  the bottom-sbottom and charged scalar loops, while
  $m_{\nu_2}^{exact}$ is the numerical result for all loops. In
  case of LMA the approximation works always better than 10 \%. For
  the LOW solution the typical error is of the order of 10 \%, while
  in extreme cases errors up to 25 \% can be found.}
\label{fig:Stau}
\end{figure}

In supersymmetric models with mSUGRA-like boundary conditions the
bottom-sbottom and the charged scalar loop usually give the most
important contribution to the neutrino mass matrix. This is
demonstrated in Fig.~\ref{fig:Stau} (left) for the set Ntrl and in
Fig.~\ref{fig:Stau} (right) for the set Stau. 
In both figures we show the ratio of the approximate-over-exact solar
neutrino mass parameter $m_{\nu_2}^{Appr}/m_{\nu_2}^{exact}$ versus
$\Dms$ in eV$^2$, where $m_{\nu_2}^{Appr}$ is the approximate loop
calculation involving the bottom-sbottom and the charged scalar loop,
while $m_{\nu_2}^{exact}$ is the exact numerical computation taking
into account all loops.

In the region of $\Dms$ appropriate for the currently preferred LMA-MSW
solution to the solar neutrino problem one finds that the approximate
calculation reproduces the exact result better than 10 \%. Only in the
set Ntrl one finds larger deviations, up to 25 \% in extreme cases,
when $\Dms$ lies in the LOW region, strongly disfavored by KamLAND.
This is due to the larger errors in the bottom-sbottom calculation in
this set for small neutrino masses as discussed above.

\section{Simplified approximation formulas}
\label{sec:simpl-appr-form}

\subsection{The solar mass}
\label{sec:solar-mass}

First we note that for nearly all points in our random sets we find
that $m_{\nu_2} \ll m_{\nu_3}$.  In other words, bilinear R-parity
breaking favors a hierarchical neutrino spectrum. Moreover, we have found
numerically that the terms proportional to ${\tilde \epsilon_i}\times
{\tilde \epsilon_j}$ in the self energies in Eq.~(\ref{eq:bsb}) give
the most important contribution to $m_{\nu_2}$ in the bottom-sbottom 
loop calculation in most points of our sets. If these terms are dominant 
one can find a very simple approximation for the bottom-sbottom loop
contribution to $m_{\nu_2}$.  It is given by
\begin{equation}
\label{Simplest}
m_{\nu_2} \simeq \frac{3}{16 \pi^2} \sin(2\theta_{\tilde b}) 
m_b \Delta B_0^{\tilde\tau_2\tilde\tau_1}\ 
\frac{({\tilde \epsilon}_1^2 + {\tilde \epsilon}_2^2)}{\mu^2}
\end{equation}

We have checked numerically that Eq.~(\ref{Simplest}) reproduces 
the result of the full approximative formula to high 
accuracy if $m_{\nu_2} \le 0.3 m_{\nu_3}$ . Note also that 
Eq. (\ref{Simplest}) holds only if the 1-loop contributions to 
the neutrino mass matrix are smaller than the tree-level one. 
This condition requires that $|{\vec \epsilon}|^2/|\Lambda| \le 1$ 
approximately, i.e. the bilinear parameters $\epsilon_i$ must 
be suppressed with respect to $\mu$. Note that such a suppression 
could, in principle, be motivated by suitable flavour 
symmetries \cite{Mira:2000gg}.

Due to the more complicated structure of the charged scalar loop it is
not possible to give a simple equation for $m_{\nu_2}$ similar to
Eq.~(\ref{Simplest}) for the bottom-sbottom loop.  However, for
$m_{\nu_2}$ larger than (few) $\times 10^{-4}$ we have found that the
most important contributions to the charged scalar loop are the
terms proportional to $\Delta B_0^{\tilde\tau_2\tilde\tau_1}$, 
$\Delta B_0^{H^{\pm}\tilde\tau_1}$ and $\Delta B_0^{H^{\pm}\tilde\tau_2}$ 
in Eq. (\ref{ChargedHiggsLoop}). We note in passing that Eq.  (\ref{Simplest}),
with appropriate replacements, allows us to estimate the typical
contributions to the charged scalar loop within a factor of $\sim 3$.
However, such an estimate will be biased toward too small (large)
$m_{\nu_2}$ for scalar tau (neutralino) LSPs.

\begin{figure}[htbp]
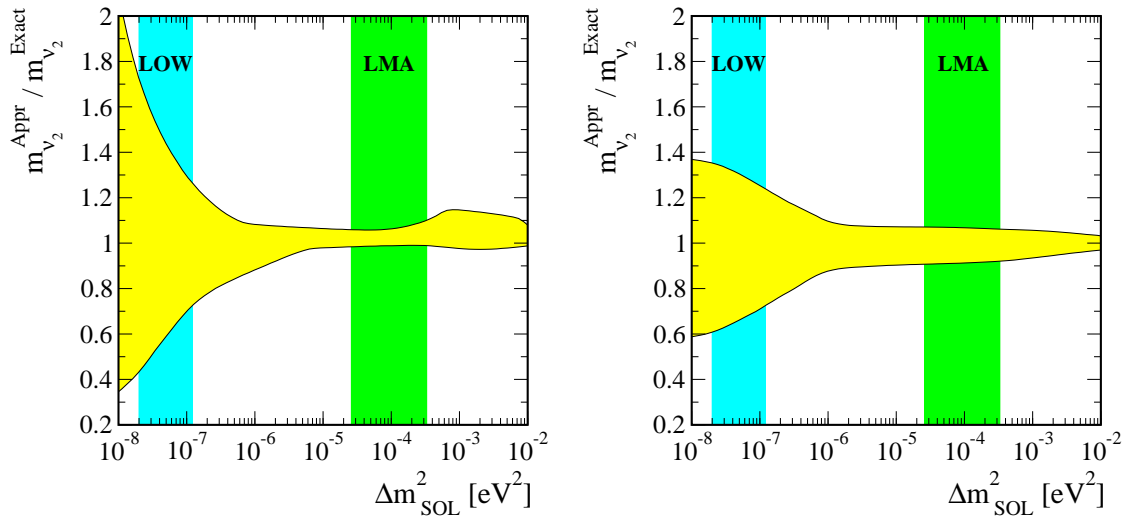
 \centering
  \begin{tabular}{cc}
    \includegraphics[width=0.45\linewidth]{Nf2002Sim-v4.eps}
    &\includegraphics[width=0.45\linewidth]{StauSim-v4.eps}
  \end{tabular}
\caption[] {($m_{\nu_2}^{Appr}/m_{\nu_2}^{exact}$) versus 
  $\Dms$ [$eV^2$] for the set Ntrl (left) and Stau (right). Shown is
  the result of the simplified approximation formula in
  Eq.~(\ref{Simplest}) for the sbottom-bottom loop and taking into 
  account only coefficients
  $C_{H^{\pm}\tilde\tau_2}$, $C_{H^{\pm}\tilde\tau_1}$ and
  $C_{\tilde\tau_2\tilde\tau_1}$ in the charged scalar loop.}
\label{fig:Simp}
\end{figure}

In Fig. (\ref{fig:Simp}) we show a comparison of our simplified approximation 
formula, including the simple form of the sbottom-bottom loop and 
the three most important coefficients for the charged scalar loop, 
as discussed above, to the full numerical calculation including all 
loops. As one can see, even the simplified version of our formula 
works surprisingly well in the LMA-MSW regime, although the agreement 
with the full calculation is now less good for the LOW region, as 
could have been expected from the results discussed previously.

\subsection{The solar mixing angle}
\label{sec:solar}

\begin{figure}[htbp]
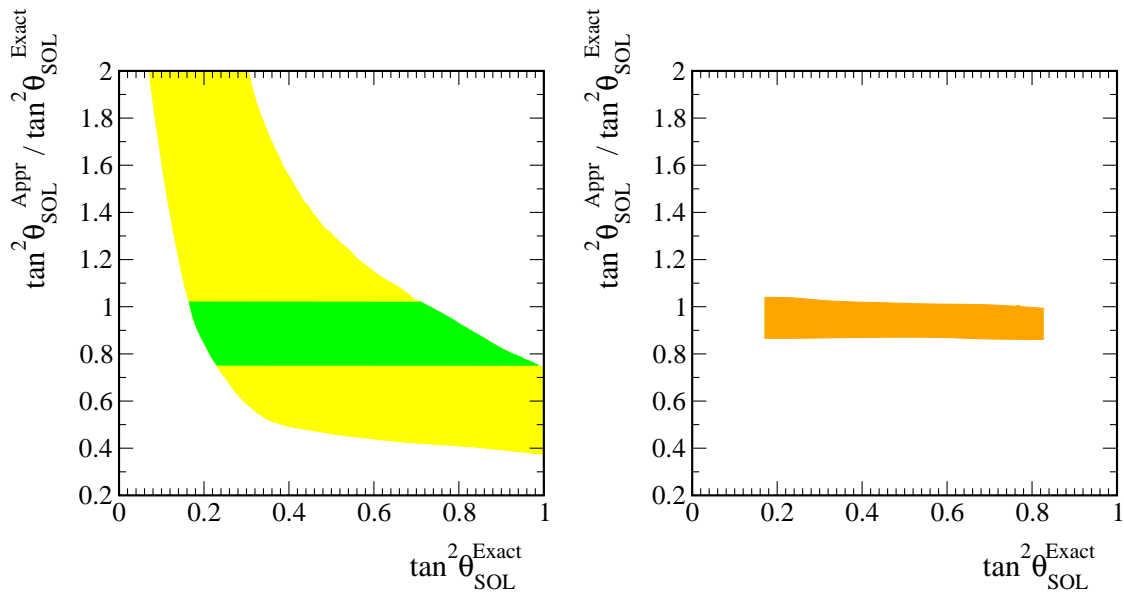
 \centering
  \begin{tabular}{cc}
    \includegraphics[width=0.45\linewidth]{Nfnd0Angle-v3.eps}
    &\includegraphics[width=0.45\linewidth]{Nfnd1Angle-v3.eps}
  \end{tabular}
\caption[] {(${ \tan^2\theta_\Sol }^{Appr}/
  {\tan^2\theta_\Sol}^{exact}$) versus ${\tan^2\theta_\Sol}^{exact}$.
  On the left panel the darker region contains over 90\% of the points
  in our sample. In the right panel the points in the region shown
  satisfy the cut $\sin(2\theta_{\tilde b}) \Delta
  B_0^{\tilde\tau_2\tilde\tau_1}> 0.02$ .}
\label{fig:SolarAngle}
\end{figure}

In the basis where the tree-level neutrino mass matrix is diagonal the
mass matrix at one--loop level can be written as 
\begin{eqnarray}
  \label{eq:2}
  \widetilde m_\nu= V_\nu^{(0)T} m_\nu V_\nu^{(0)} 
      =\left( \begin{array}{ccc}
      c_1 \widetilde \epsilon_1 \widetilde \epsilon_1 & c_1 \widetilde
      \epsilon_1 \widetilde \epsilon_2  
      & c_1 \widetilde \epsilon_1 \widetilde \epsilon_3 \\
      c_1 \widetilde \epsilon_2 \widetilde \epsilon_1 & c_1 \widetilde
      \epsilon_2 \widetilde \epsilon_2  
      & c_1 \widetilde \epsilon_2 \widetilde \epsilon_3 \\
      c_1 \widetilde \epsilon_3 \widetilde \epsilon_1 & c_1 \widetilde
      \epsilon_3 \widetilde \epsilon_2  
      & c_0 |\vec\Lambda|^2 + c_1 \widetilde \epsilon_3 \widetilde \epsilon_3 
     \end{array} \right) + \cdots
\end{eqnarray}
where the $\widetilde \epsilon_i$ were defined before in
Eq.~(\ref{eq:TildeEpsilon}). Coefficients $c_0$ and $c_1$ contain 
couplings and supersymmetric masses. Since they cancel in the final 
expression for the angle their exact definition is not necessary in 
the following. Dots stand for other terms which we will 
assume to be less important in the following, see the discussion at 
the end of this subsection.  
This matrix can be diagonalized approximately under the condition 
\begin{eqnarray}
  \label{eq:3}
  x\equiv\frac{c_1 |\vec{\widetilde{\epsilon}}|^2}{c_0 |\vec \Lambda|^2 }\ll 1
\end{eqnarray}
i.e. if the 1-loop contribution to the neutrino mass matrix is smaller 
than the tree-level contribution, as also discussed above for 
Eq. (\ref{Simplest}). Then
\begin{eqnarray}
  \label{eq:5}
  \widetilde m_\nu= c_0 |\vec\Lambda|^2 
      \left( \begin{array}{ccc} \displaystyle
      x\, \frac{\widetilde \epsilon_1 \widetilde \epsilon_1 }{
      |\vec{\widetilde \epsilon|^2} } 
      &\displaystyle x\, \frac{\widetilde \epsilon_1 \widetilde
      \epsilon_2 }{|\vec{\widetilde \epsilon|^2}} 
      & \displaystyle x\, \frac{\widetilde \epsilon_1 \widetilde
      \epsilon_3 }{|\vec{\widetilde \epsilon|^2}} \\[+4mm]
      \displaystyle
      x\, \frac{\widetilde \epsilon_2 \widetilde \epsilon_1 }{
      |\vec{\widetilde \epsilon|^2 }}
      &\displaystyle x\, \frac{\widetilde \epsilon_2 \widetilde
      \epsilon_2 }{|\vec{\widetilde \epsilon|^2}}
      & \displaystyle x\, \frac{\widetilde \epsilon_2 \widetilde
      \epsilon_3 }{|\vec{\widetilde \epsilon|^2}} \\[+4mm] 
      \displaystyle
      x\, \frac{\widetilde \epsilon_3 \widetilde \epsilon_1 }{
        |\vec{\widetilde \epsilon|^2 }}
      &\displaystyle x\, \frac{\widetilde \epsilon_3 \widetilde
      \epsilon_2 }{|\vec{\widetilde \epsilon|^2}}
      & \displaystyle 1 + x\, \frac{\widetilde \epsilon_3 \widetilde
      \epsilon_3 }{|\vec{\widetilde \epsilon|^2}}  
     \end{array} \right)
\end{eqnarray}
We now calculate the eigenvalues and eigenvectors of this matrix as
series expansions in the small $x$ parameter. For the eigenvalues we
get
\begin{eqnarray}
  \label{eq:6}
  m_1&=&0\nonumber  \\
  m_2&=& x\, c_0 \frac{|\vec\Lambda|^2}{|\vec{\widetilde \epsilon|^2}}
 + \mathcal{O}(x^2)  = 
  c_1 \left( \widetilde \epsilon_1^2+\widetilde \epsilon_2^2 \right)
  + \mathcal{O}(x^2)\\
  m_3&=& c_0 |\vec\Lambda|^2 + c_1 \widetilde \epsilon_3^2 +
  \mathcal{O}(x^2) \nonumber 
\end{eqnarray}
and for the first two eigenvalues (the third can also be easily
obtained but it will not be necessary for the discussion of the solar
mixing angle),
\begin{eqnarray}
  \label{eq:7}
  e_1&=&\left(-\frac{\widetilde \epsilon_2}{\widetilde \epsilon_1}\, 
  \sqrt{\frac{\widetilde \epsilon_1^2}{\widetilde \epsilon_1^2+
   \widetilde \epsilon_2^2}},\ \sqrt{\frac{\widetilde
  \epsilon_1^2}{\widetilde \epsilon_1^2+ \widetilde
  \epsilon_2^2}},\ 0\right)\nonumber \\[+2mm]
  e_2&=&\left( e_{2,1}, e_{2,2}, e_{2,3}\right)
\end{eqnarray}
where up to $\mathcal{O}(x^2)$ we have,
\begin{eqnarray}
  \label{eq:8}
  e_{2,1}&=& -\frac{\widetilde \epsilon_1 \widetilde \epsilon_3}
  {\sqrt{\widetilde \epsilon_3^2(\widetilde \epsilon_1^2 +\widetilde
  \epsilon_2^2 )}} +\frac{1}{2}\, \frac{\widetilde \epsilon_1
  \widetilde \epsilon_3 \sqrt{\widetilde \epsilon_3^2(\widetilde
  \epsilon_1^2 +\widetilde \epsilon_2^2 )}}{|\vec{\widetilde
  \epsilon|^4}}\ x^2 + \mathcal{O}(x^3)\nonumber
 \\[+2mm]
  e_{2,2}&=& -\frac{\widetilde \epsilon_2 \widetilde \epsilon_3}
  {\sqrt{\widetilde \epsilon_3^2(\widetilde \epsilon_1^2 +\widetilde
  \epsilon_2^2 )}}+\frac{1}{2}\, \frac{\widetilde \epsilon_2
  \widetilde \epsilon_3 \sqrt{\widetilde \epsilon_3^2(\widetilde
  \epsilon_1^2 +\widetilde \epsilon_2^2 )}}{|\vec{\widetilde
  \epsilon|^4}}\ x^2 + \mathcal{O}(x^3)
 \\[+2mm]
  e_{2,3}&=& \frac{\sqrt{\widetilde \epsilon_3^2(\widetilde
  \epsilon_1^2 +\widetilde  \epsilon_2^2 )}}{|\vec{\widetilde
  \epsilon|^2}}\ x + \frac{\left(\widetilde  \epsilon_1^2 +\widetilde
  \epsilon_2^2-\widetilde \epsilon_3^2 \right) \sqrt{\widetilde
  \epsilon_3^2(\widetilde \epsilon_1^2 +\widetilde
  \epsilon_2^2)}}{|\vec{\widetilde  \epsilon|^4}}\ x^2 +
  \mathcal{O}(x^3) \nonumber 
\end{eqnarray}
Knowing the eigenvectors we can write down the rotation matrix that
diagonalizes $\widetilde m_\nu$,
\begin{eqnarray}
  \label{eq:9}
  \widetilde V_\nu^T \widetilde m_\nu \widetilde V_\nu
  =\hbox{diag}(m_1,m_2,m_3) 
\end{eqnarray}
where
\begin{eqnarray}
  \label{eq:10}
   \widetilde V_\nu^T= \left( \begin{array}{ccc}
       e_{1,1} & e_{1,2}  & e_{1,3} \\
       e_{2,1} & e_{2,2}  & e_{2,3} \\
       e_{3,1} & e_{3,2}  & e_{3,3} 
       \end{array}\right)
\end{eqnarray}
The neutrino mixing matrix is then given by
\begin{eqnarray}
  \label{eq:11}
  U=\left(  V_\nu^T  \widetilde V_\nu^T\right)^T
\end{eqnarray}
Using the fact that $U_{e3}$ has to be small one can get the following
expression for the solar mixing angle:
\begin{eqnarray}
  \label{eq:sol1}
  \tan^2\theta_\Sol  = \frac{U_{e2}^2}{U_{e1}^2}
\end{eqnarray}
Now using the Eqs.~(\ref{eq:10}), (\ref{eq:8}) and substituting in
Eq.~(\ref{eq:11}) we obtain the very simple expression for the solar
mixing angle,
\begin{eqnarray}
  \label{eq:12}
   \tan^2\theta_\Sol  = \frac{\widetilde  \epsilon_1^2}{\widetilde
  \epsilon_2^2} 
\end{eqnarray}
This formula is a very good approximation if the one--loop matrix has 
the structure $\epsilon_i \times \epsilon_j$, as is the case of the
bottom-sbottom loop (and, to a lesser extend also for the charged scalar 
loop, which has one coefficient with the same index structure), and if 
$m_{\nu_3} \gg m_{\nu_2}$ . This is illustrated in Fig.~\ref{fig:SolarAngle}. 

In the left panel we show a calculation comparing for all points in the 
set Ntrl the approximate to the exact solar angle, while the 
right panel shows a subset of points using the cut 
$\sin(2\theta_{\tilde b}) \Delta B_0^{\tilde\tau_2\tilde\tau_1}>0.02$. 
Note that this cut is designed such as to prefer points in which 
there is a sizeable contribution to the full 1-loop neutrino mass 
due to the bottom-sbottom loop. For points in which the charged 
scalar loop dominates eq. (\ref{eq:12}) gives only a factor-of-two 
estimate of the true solar angle.

Note finally that eq. (\ref{eq:12}) will fail completely, if 
$\Lambda_{\mu} \equiv \Lambda_{\tau}$ and $\epsilon_{\mu} \equiv 
\epsilon_{\tau}$, since then ${\widetilde \epsilon_2^2} =0$, see 
Eq. (\ref{eq:TildeEpsilon}). This is the origin of the ``sign 
condition'' discussed in \cite{Hirsch:2000ef}. 

\section{Discussion and conclusions}
\label{sec:disc-concl}

We have presented an approximate calculation of the neutrino mass
matrix at one-loop in supersymmetry with bilinearly broken R-parity.
The method is based on a systematic perturbative expansion of R-parity
violating vertices to leading order.  We have identified the
bottom-sbottom and the charged scalar loop as the most important ones,
at least in supersymmetric models with mSUGRA-like boundary
conditions.  Taking into account only these loops, we have given
explicit formulas and discussed their validity as well as the accuracy
with which they describe solar neutrino mass and mixing parameters.
This was done by comparing our analytical results to the exact
numerical calculation.  We have found that for the case of the large
mixing MSW solution our formulas - even within the simplified form 
Eq. (\ref{Simplest}) and Eq. (\ref{eq:12}) -  yield good agreement 
with the full
numerical calculation, but are much simpler to implement than the full
numerical one-loop calculation.  The only solar neutrino ``solutions''
for which our analytical approximation is less accurate are those that 
are now ruled out by the recent reactor neutrino data from KamLAND.  

Let us finally discuss some possible caveats to the success of our 
approximate treatment. One is the assumption that supersymmetry breaking 
mass terms are flavour diagonal, which we have adopted, motivated by 
constraints from flavour changing processes. Although such terms could 
be included into our approximate treatment, we have not done so, mainly 
due to the fact that the resulting formulas would be much more complicated 
and, therefore, of very limited practical use. A second concern is 
that our sample points were all generated using mSugra assumptions 
for the soft breaking masses. Clearly there are other possibilities 
to break supersymmetry and even though we expect that the bottom-sbottom 
loop and the charged scalar loop will still be well described by our 
approximation formulas, other loops, which we didn't take into account, 
might be more important than what we have found in our data sets.

\section*{Acknowledgements}  

This work was supported by Spanish grant BFM2002-00345, by the
European Commission RTN grant HPRN-CT-2000-00148 and the ESF
\emph{Neutrino Astrophysics Network}.  M. H.  is supported by a
Spanish MCyT Ramon y Cajal contract. W.~P.~is supported by the 'Erwin
Schr\"odinger fellowship No.~J2095' of the `Fonds zur F\"orderung der
wissenschaftlichen Forschung' of Austria FWF and partly by the Swiss
`Nationalfonds'. M.A.D. was supported by Conicyt grant No. 1010974. 
J.R. was partly supported by the fellowship SFRH/BSAB/269/2002 from FCT
(Portugal) with funds from the EU. 

\newpage

\appendix

\section{Rotation Matrices}
\label{ap:Rotations}


In the basis $(H_d^+, H_u^+, \widetilde{e}_L^+, \widetilde{\mu}_L^+,
\widetilde{\tau}_L^+,\widetilde{e}_R^+, \widetilde{\mu}_R^+, 
\widetilde{\tau}_R^+)$, one can write, to first order in R-parity 
violating parameters, the Goldstone rotation matrix as
\begin{equation}
{\bf R}_G=\left[\matrix{
c_{\beta} & -s_{\beta} & v_1/v & v_2/v & v_3/v & 0 & 0 & 0 \cr
s_{\beta} & c_{\beta} & 0 & 0 & 0 & 0 & 0 & 0 \cr
-c_{\beta}v_1/v & s_{\beta}v_1/v & 1 & 0 & 0 & 0 & 0 & 0 \cr
-c_{\beta}v_2/v & s_{\beta}v_2/v & 0 & 1 & 0 & 0 & 0 & 0 \cr
-c_{\beta}v_3/v & s_{\beta}v_3/v & 0 & 0 & 1 & 0 & 0 & 0 \cr
0 & 0 & 0 & 0 & 0 & 1 & 0 & 0 \cr
0 & 0 & 0 & 0 & 0 & 0 & 1 & 0 \cr
0 & 0 & 0 & 0 & 0 & 0 & 0 & 1 \cr
}\right]
\label{eq:GoldstoneRotation}
\end{equation}
where $v^2=v_d^2+v_u^2$ to this order, and $\tan\beta=v_u/v_d$, as
usual. We have also used the shorthand notation
$c_{\beta}(s_{\beta})=\cos\beta (\sin\beta)$.


Neglecting the electron and muon Yukawa couplings, the rotation that
diagonalizes the sleptons at tree level is given by (in the same basis
as above)
\begin{equation}
{\bf R}_{\tilde\tau}=\left[\matrix{
1 & 0 & 0 & 0 & 0 & 0 & 0 & 0 \cr
0 & 1 & 0 & 0 & 0 & 0 & 0 & 0 \cr
0 & 0 & 1 & 0 & 0 & 0 & 0 & 0 \cr
0 & 0 & 0 & 1 & 0 & 0 & 0 & 0 \cr
0 & 0 & 0 & 0 & c_{\tilde\tau} & 0 & 0 & s_{\tilde\tau} \cr
0 & 0 & 0 & 0 & 0 & 1 & 0 & 0 \cr
0 & 0 & 0 & 0 & 0 & 0 & 1 & 0 \cr
0 & 0 & 0 & 0 & -s_{\tilde\tau} & 0 & 0 & c_{\tilde\tau} \cr
}\right]
\label{eq:StauRotation}
\end{equation}
%


After the rotations ${\bf R}_{\tilde\tau}{\bf R}_G$ are performed, the 
charged scalar mass matrix is diagonalized up to small R-parity 
violating entries. In the approximation where there is no intergenerational 
mixing and $h_{\mu}\approx h_e\approx0$, these are

\begin{equation}
{\bf \Delta M^2_{S^{\pm}}}_{\tilde\tau}=\left[\matrix{
0 & 0 & 0 & 0 & 0 & 0 & 0 & 0 \cr
0 & 0 & \widetilde X_{HL_1} & \widetilde X_{HL_2} & \widetilde
X_{HL_3} & 0 & 0 & 
\widetilde X_{HR_3}\cr 
0 & \widetilde X_{HL_1} & 0 & 0 & 0 & 0 & 0 & 0 \cr
0 & \widetilde X_{HL_2} & 0 & 0 & 0 & 0 & 0 & 0 \cr
0 & \widetilde X_{HL_3} & 0 & 0 & 0 & 0 & 0 & 0 \cr
0 & 0 & 0 & 0 & 0 & 0 & 0 & 0 \cr
0 & 0 & 0 & 0 & 0 & 0 & 0 & 0 \cr
0 & \widetilde X_{HR_3} & 0 & 0 & 0 & 0 & 0 & 0 \cr
}\right]
\label{eq:OneLoopMixings}
\end{equation}
where
\begin{equation}
\begin{tabular}{ll}
$\displaystyle 
\widetilde{X}_ {HL_i}=X_{HL_i}$
&
$\displaystyle 
\widetilde{X}_{HR_i}=X_{HR_i}, \ (i=1,2)$
\\[+2mm]
$\displaystyle 
\widetilde{X}_{HL_3}=c_{\tilde\tau} X_{uL_3}+s_{\tilde\tau}X_{dR_3}$
&
$\displaystyle 
\widetilde{X}_{HR_3}=-s_{\tilde\tau} X_{HL_3}+c_{\tilde\tau}X_{HR_3}$
\end{tabular}
\end{equation}
with
\begin{equation}
\begin{tabular}{ll}
$\displaystyle X_{HL_i}=s_{\beta}X_{uL_i}+c_{\beta}X_{dL_i}$&
$\displaystyle  X_{HR_3}=s_{\beta}X_{uR_3}+c_{\beta}X_{dR_3}$, \
(i=1,3)
\end{tabular}
\end{equation}
and
\begin{equation}
\begin{tabular}{ll}
$X_{uL_i}=\textstyle{1\over4}g^2v_dv_i-\mu\epsilon_i-\textstyle{1\over2}
h_{\tau}^2v_dv_i \delta_{i3}\,,$
&
$X_{dL_i}={{v_i}\over{v_d}}{{c_{\beta}}\over{s_{\beta}}}
m_{\tilde\nu}^2-\mu\epsilon_i{{c_{\beta}}\over{s_{\beta}}}
+\textstyle{1\over4}g^2v_uv_i\,,$
\\[+2mm]
$X_{uR_3}=-\textstyle{1\over{\sqrt{2}}}h_{\tau}(A_{\tau}v_3+\epsilon_3v_u)\,,
$&
$X_{dR_3}=-\textstyle{1\over{\sqrt{2}}}h_{\tau}(\mu v_3+\epsilon_3v_d)\,.$
\end{tabular}
\end{equation}
These mixings are removed with the rotation matrix ${\bf R_X}$ given
by 
\begin{equation}
{\bf R_X}=\left[\matrix{
1 & 0 & 0 & 0 & 0 & 0 & 0 & 0 \cr
0 & 1 &  \Theta_{HL_1} & \Theta_{HL_2} & \Theta_{HL_3} & 0 & 0 & 
\Theta_{HR_3} \cr 
0 & -\Theta_{HL_1} & 1 & 0 & 0 & 0 & 0 & 0 \cr
0 & -\Theta_{HL_2} & 0 & 1 & 0 & 0 & 0 & 0 \cr
0 & -\Theta_{HL_3} & 0 & 0 & 1 & 0 & 0 & 0 \cr
0 & 0 & 0 & 0 & 0 & 1 & 0 & 0 \cr
0 & 0 & 0 & 0 & 0 & 0 & 1 & 0 \cr
0 & -\Theta_{HR_3} & 0 & 0 & 0 & 0 & 0 & 1 \cr
}\right]
\label{eq:XRotation}
\end{equation}
in the small mixing approximation $\sin \Theta \simeq \Theta$. Note
that here we have defined
\begin{equation}
\Theta_{HL_i}\equiv
\frac{\widetilde X_{HL_i}}
{m^2_{H^{\pm}} - m^2_{\tilde \ell_{Li}}}, \quad
\Theta_{HR_i}\equiv
\frac{\widetilde X_{HR_i}}
{m^2_{H^{\pm}} - m^2_{\tilde \ell_{Ri}}}
\end{equation}
%


Putting everything together we get the final form of the charged
scalar diagonalization matrix ${\bf R}_X{\bf R}_{\tilde\tau}{\bf R}_G$
which can be expressed as

\begin{eqnarray}
\label{eq:FinalRotation}
{\bf R}_X{\bf R}_{\tilde\tau}{\bf R}_G&=&\\
&&\nonumber\\
&&\!\!\!\!\!\!\!\!\!\!\!\!\!\!\!\!\!\!\!\!\!\!\!\!\!\!\!\!\!\!\!\!\!\!\!\!
\left[\matrix{
c_{\beta} & -s_{\beta} & v_1/v & v_2/v & v_3/v & 0 & 0 & 0 \cr
s_{\beta} & c_{\beta} & \Theta_{HL_1} & \Theta_{HL_2} &\widetilde 
\Theta_{HL_3} &
0 & 0 & \widetilde \Theta_{HR_3} \cr
- s_{\beta}\Theta_{HL_1} - c_{\beta}{{v_1}\over v} & 
- c_{\beta}\Theta_{HL_1} + s_{\beta}{{v_1}\over v} & 
1 & 0 & 0 & 0 & 0 & 0 \cr
- s_{\beta}\Theta_{HL_2}-c_{\beta}{{v_2}\over v} & 
- c_{\beta}\Theta_{HL_2}+s_{\beta}{{v_2}\over v} & 
0 & 1 & 0 & 0 & 0 & 0 \cr
-s_{\beta}\Theta_{HL_3}-c_{\tilde\tau}c_{\beta}{{v_3}\over v} & 
-c_{\beta}\Theta_{HL_3}+c_{\tilde\tau}s_{\beta}{{v_3}\over v} & 
0 & 0 & c_{\tilde\tau} & 0 & 0 & s_{\tilde\tau} \cr
0 & 0 & 0 & 0 & 0 & 1 & 0 & 0 \cr
0 & 0 & 0 & 0 & 0 & 0 & 1 & 0 \cr
-s_{\beta}\Theta_{HR_3}+s_{\tilde\tau}c_{\beta}{{v_3}\over v} & 
-c_{\beta}\Theta_{HR_3}-s_{\tilde\tau}s_{\beta}{{v_3}\over v} & 
0 & 0 & -s_{\tilde\tau} & 0 & 0 & c_{\tilde\tau}
}\right]
\nonumber
\end{eqnarray}
where we have defined,
\begin{equation}
  \widetilde\Theta_{HL_3}=c_{\tilde\tau} \Theta_{HL_3}
- s_{\tilde\tau} \Theta_{HR_3}\ , \
  \widetilde\Theta_{HR_3}=s_{\tilde\tau} \Theta_{HL_3}
+ c_{\tilde\tau} \Theta_{HR_3}\ . \
\end{equation}

\newpage
\section{Charged Higgs/slepton couplings}
\label{ap:couplings}

The couplings of the five (generalized to include also the three
charged leptons) charginos to the eight charged scalars (including
Higgs bosons and sleptons of both chiralities) and seven neutralinos
(generalized to include also the three neutrinos) are given
by~\cite{Hirsch:2000ef}
\begin{eqnarray}
O^{cns}_{Lijk}&=&R^{S^{\pm}}_{k1}h_{\tau}{\cal N}_{j7}{\cal V}_{i5}
-R^{S^{\pm}}_{k2}\left({g\over{\sqrt{2}}}{\cal N}_{j2}{\cal V}_{i2}+
{g'\over{\sqrt{2}}}{\cal N}_{j1}{\cal V}_{i2}+g{\cal N}_{j4}{\cal V}_{i1}
\right)
\\&&
-R^{S^{\pm}}_{k5}h_{\tau}{\cal N}_{j3}{\cal V}_{i5}
-g'\sqrt{2}\left(R^{S^{\pm}}_{k6}{\cal N}_{j1}{\cal V}_{i3}
+R^{S^{\pm}}_{k7}{\cal N}_{j1}{\cal V}_{i4}
+R^{S^{\pm}}_{k8}{\cal N}_{j1}{\cal V}_{i5}\right)
\nonumber
\label{eq:ocnsL}
\end{eqnarray}
where $i$ labels the charginos, $j$ labels neutralinos, and $k$ labels the
charged scalars, respectively. For the the right-handed couplings the
corresponding couplings are given by
\begin{eqnarray}
O^{cns}_{Rijk}&=&R^{S^{\pm}}_{k1}\left({g\over{\sqrt{2}}}{\cal N}_{j2}\,
{\cal U}_{i2}
+{g'\over{\sqrt{2}}}{\cal N}_{j1}\,{\cal U}_{i2}-g{\cal N}_{j3}\,
{\cal U}_{i1}\right)
\nonumber\\&&
+R^{S^{\pm}}_{k3}\left({g\over{\sqrt{2}}}{\cal N}_{j2}\,{\cal U}_{i3}
+{g'\over{\sqrt{2}}}{\cal N}_{j1}\,{\cal U}_{i3}-
g{\cal N}_{j5}\,{\cal U}_{i1}\right)
\nonumber\\&&
+R^{S^{\pm}}_{k4}\left({g\over{\sqrt{2}}}{\cal N}_{j2}\,{\cal U}_{i4}
+{g'\over{\sqrt{2}}}{\cal N}_{j1}\,{\cal U}_{i4}-
g{\cal N}_{j6}\,{\cal U}_{i1}\right)
\\&&
+R^{S^{\pm}}_{k5}\left({g\over{\sqrt{2}}}{\cal N}_{j2}\,{\cal U}_{i5}
+{g'\over{\sqrt{2}}}{\cal N}_{j1}\,{\cal U}_{i5}-
g{\cal N}_{j7}\,{\cal U}_{i1}\right)
\nonumber\\&&
+R^{S^{\pm}}_{k8}h_{\tau}\left({\cal N}_{j7}\,{\cal U}_{i2}-
{\cal N}_{j3}\,{\cal U}_{i5}\right)
\nonumber
\label{eq:ocnsR}
\end{eqnarray}

After approximating the rotation matrices ${\cal U}$, and ${\cal V}$
in the chargino sector, and ${\cal N}$ in the neutralino sector we
find the expressions given in Eqs.~
(\ref{eq:charginoL})-(\ref{eq:leptonR}).  Note that we have divided
them into cases where the charged fermion is a lepton or a chargino. For
the left couplings when the charged fermion is a chargino we have,
\begin{equation}
O^{cns}_{Lijk}=R^{S^{\pm}}_{k2}\left[{g\over{\sqrt{2}}}a_2V_{i'2}
+{g'\over{\sqrt{2}}}a_1V_{i'2}+ga_4V_{i'1}\right]
|\vec\Lambda|\delta_{j3}
\label{eq:charginoL}
\end{equation}
where $V$ is the reduced $2\times2$ chargino diagonalization matrix of the
MSSM, and $i'=1,2$. If the charged fermion is a lepton we have
\begin{eqnarray}
O^{cns}_{Lijk}&=&R^{S^{\pm}}_{k1}h_{\tau}V^T_{\nu,j3}\delta_{i3}
+R^{S^{\pm}}_{k5}h_{\tau}\left(b\tilde\epsilon_j+
a_3|\vec\Lambda|\delta_{j3}\right)\delta_{i3}
\nonumber\\&&
+\left[R^{S^{\pm}}_{k6}\delta_{i1}+R^{S^{\pm}}_{k7}\delta_{i2}
+R^{S^{\pm}}_{k8}\delta_{i3}\right]\sqrt{2}g'a_1|\vec\Lambda|\delta_{j3}
\label{eq:leptonL}
\end{eqnarray}
For the right-handed couplings when the charged fermion is a chargino
we get
\begin{eqnarray}
O^{cns}_{Rijk}&=&R^{S^{\pm}}_{k1}\left[
-{\textstyle{1\over{\sqrt{2}}}}(ga_2+g'a_1)|\vec\Lambda|
\delta_{j3}U_{i'2}
+g(b\tilde\epsilon_j+a_3|\vec\Lambda|\delta_{j3})U_{i'1}\right]
\nonumber\\&&
-R^{S^{\pm}}_{k3}gV^T_{\nu,j1}U_{i'1}-R^{S^{\pm}}_{k4}
gV^T_{\nu,j2}U_{i'1}
-R^{S^{\pm}}_{k5}gV^T_{\nu,j3}U_{i'1}
+R^{S^{\pm}}_{k8}h_{\tau}V^T_{\nu,j3}U_{i'2}
\label{eq:charginoR}
\end{eqnarray}
where $U$ is the second $2\times2$ chargino rotation matrix of the MSSM.
Finally, if the charged fermion is a lepton one has
\begin{eqnarray}
\label{eq:leptonR}
O^{cns}_{Rijk}&=&
-R^{S^{\pm}}_{k3}\left[
{\textstyle{1\over{\sqrt{2}}}}(ga_2+g'a_1)|\vec\Lambda|
\delta_{j3}\delta_{i1}-gV^T_{\nu,j1}a_1^L\Lambda_i\right]
\nonumber\\&&
-R^{S^{\pm}}_{k4}\left[
{\textstyle{1\over{\sqrt{2}}}}(ga_2+g'a_1)|\vec\Lambda|
\delta_{j3}\delta_{i2}-gV^T_{\nu,j2}a_1^L\Lambda_i\right]
\\&&
-R^{S^{\pm}}_{k5}\left[
{\textstyle{1\over{\sqrt{2}}}}(ga_2+g'a_1)|\vec\Lambda|
\delta_{j3}\delta_{i3}-gV^T_{\nu,j3}a_1^L\Lambda_i\right]
\nonumber\\&&
-R^{S^{\pm}}_{k8}h_{\tau}\left[
V^T_{\nu,j3}(a_2^L\Lambda_i+b\epsilon_i)-
(b\tilde\epsilon_j+a_3|\vec\Lambda|\delta_{j3})\delta_{i3}\right]
\nonumber
\end{eqnarray}

\newpage

\section{Charged Scalar--Charged Fermion Loops}
\label{ap:diagrams}

There are nine different terms contributing to the Charged
Scalar-Charged Fermion loop, as it was shown in
Eq.~(\ref{ChargedHiggsLoop}). All these terms give a finite 
contribution to the $2\times2$ submatrix corresponding to the
light neutrinos. In this Appendix we will explain with graphs the
origin of the different terms. The conventions used were explained in
section~\ref{sec:ChargedScalarLoops}. 

\subsection{$\Delta B_0^{\tilde\tau_2\tilde\tau_1}$}

The terms proportional to $\Delta B_0^{\tilde\tau_2\tilde\tau_1}$ come
from the graphs of Fig.~\ref{fig:ChargedScalarLoop} as explained in
section~\ref{sec:ChargedScalarLoops}.

\subsection{$\Delta B_0^{H^+\tilde\tau_1}$ and  $\Delta
B_0^{H^+\tilde\tau_2}$}

\begin{figure}[htb]
\begin{center}
\begin{tabular}{cc}
\includegraphics[width=0.45\linewidth]{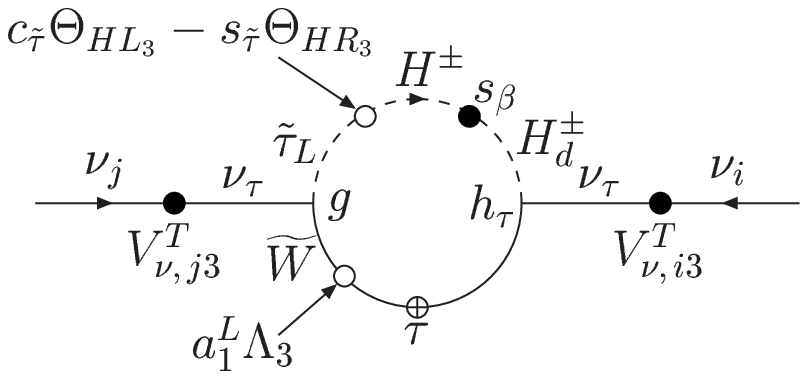}
&\includegraphics[width=0.45\linewidth]{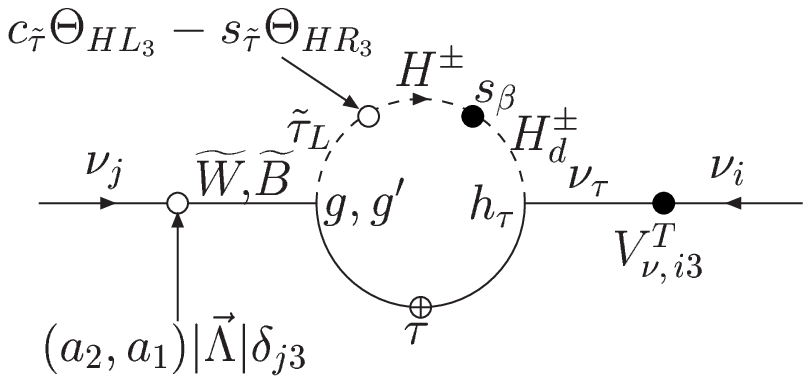}
\end{tabular}
\begin{tabular}{cc}
\includegraphics[width=0.45\linewidth]{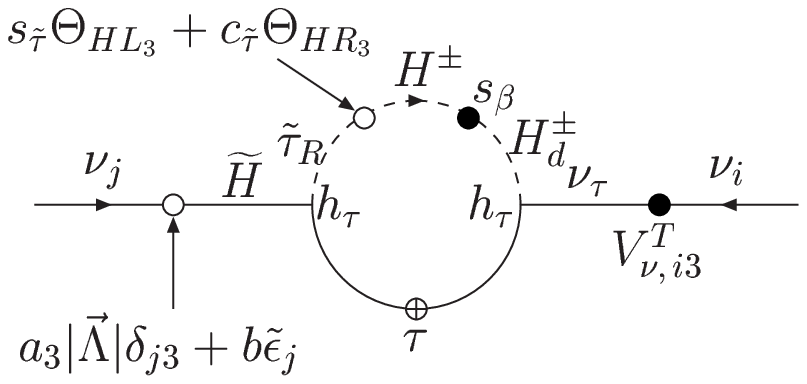}
&\includegraphics[width=0.45\linewidth]{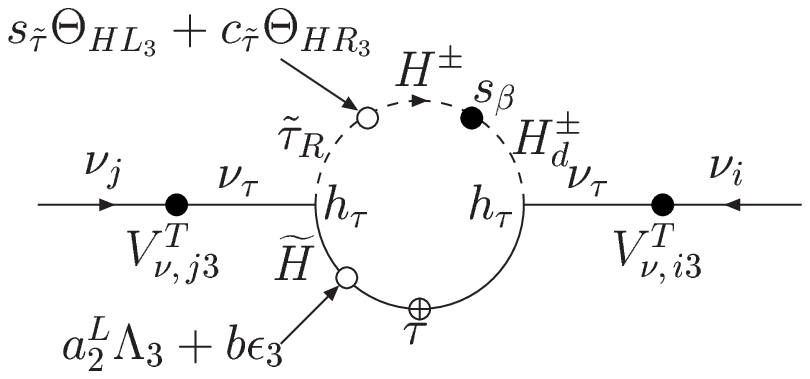}
\end{tabular}
\end{center}
\vspace{-3mm}
\caption{$H^{\pm}$ contribution to  $\Delta B_0^{H^{\pm}\tilde\tau_1}$}
\label{fig:ChargedScalarLoop-2}
\end{figure}

\begin{figure}[ht]
\begin{center}
\begin{tabular}{cc}
\includegraphics[width=0.45\linewidth]{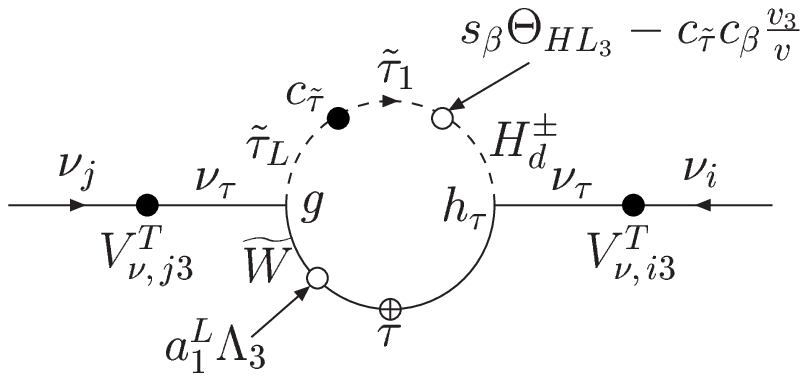}
&\includegraphics[width=0.45\linewidth]{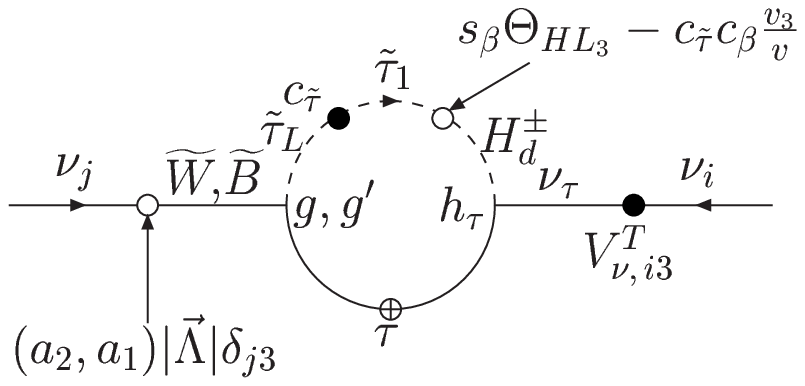}
\end{tabular}
\begin{tabular}{cc}
\includegraphics[width=0.45\linewidth]{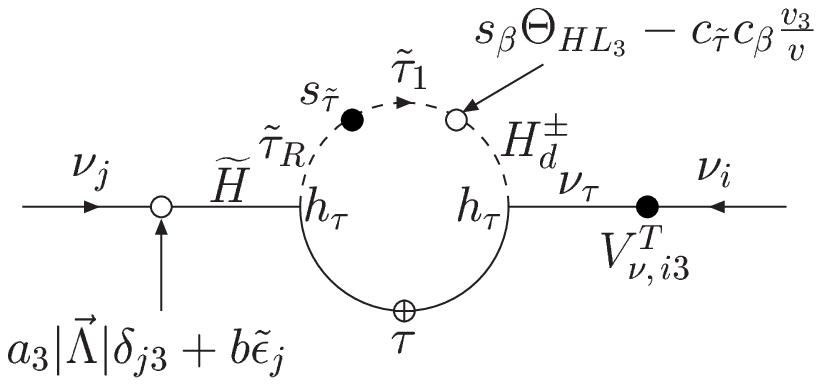}
&\includegraphics[width=0.45\linewidth]{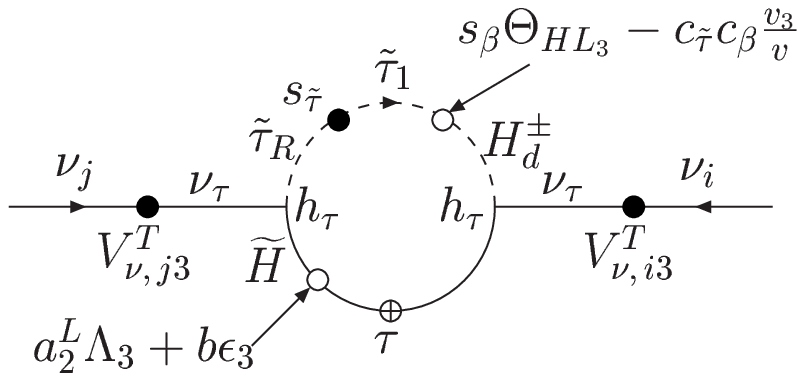}
\end{tabular}
\end{center}
\vspace{-3mm}
\caption{$\tilde\tau_1$ contributions to  
$\Delta B_0^{H^{\pm}\tilde\tau_1}$ and $\Delta
B_0^{G^{\pm}\tilde\tau_1\tilde\tau_2}$ }
\label{fig:ChargedScalarLoop-3}
\end{figure}

Now consider the terms proportional to $\Delta B_0^{H^+\tilde\tau_1}$ and
$\Delta B_0^{H^+\tilde\tau_2}$ in Eq.~(\ref{ChargedHiggsLoop}).
Of these terms, the ones which are related to the charged Higgs mixing
with staus, can be understood as coming from the four graphs of
Fig.~\ref{fig:ChargedScalarLoop-2}. 
Associated to these charged Higgs graphs are those related to the
$\tilde\tau_1$  mixing with charged Higgs. These are given in 
Fig.~\ref{fig:ChargedScalarLoop-3}.

\noindent
There is another set of four graphs corresponding to ${\tilde\tau_2}$
that are obtained from those in Fig.~\ref{fig:ChargedScalarLoop-3} by
replacing  
$\tilde\tau_1\rightarrow\tilde\tau_2$, 
$s_{\tilde\tau}\rightarrow c_{\tilde\tau}$ and
$c_{\tilde\tau}\rightarrow -s_{\tilde\tau}$. These three groups of 
four graphs, when combined, form a set which is ultraviolet
finite and account for the terms in Eq.~(\ref{ChargedHiggsLoop})
proportional to $\Delta B_0^{H^+\tilde\tau_1}$ and  $\Delta
B_0^{H^+\tilde\tau_2}$.

\subsection{$\Delta B_0^{H^{\pm}L_1}$ and  $\Delta
B_0^{H^{\pm}L_2}$}

\begin{figure}[ht]
\begin{center}
\begin{tabular}{cc}
\includegraphics[width=0.45\linewidth]{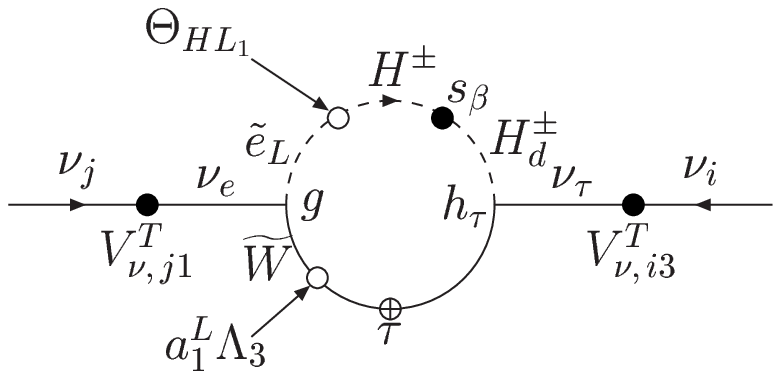}
&\includegraphics[width=0.45\linewidth]{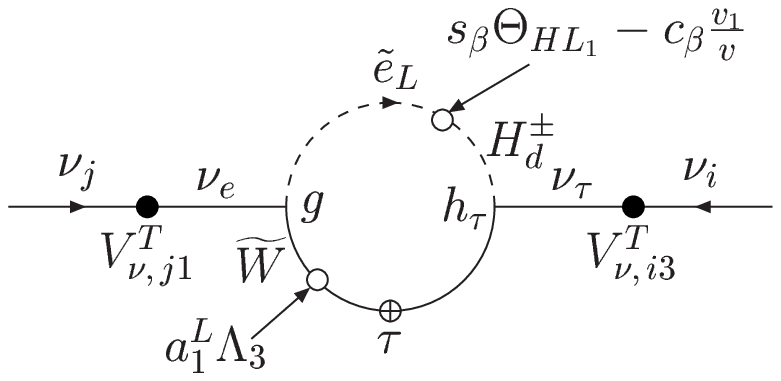}
\end{tabular}
\end{center}
\vspace{-3mm}
\caption{
a) $H^{\pm}$ contribution to  $\Delta B_0^{H^{\pm}L_1}$;
b) $\tilde{e}_L$ contribution to
$\Delta B_0^{H^{\pm}L_1}$ and $\Delta B_0^{G^{\pm}L_1}$
}
\label{fig:ChargedScalarLoop-4}
\end{figure}

We now turn our attention to the terms proportional to $\Delta
B_0^{H^{\pm}L_1}$ and $ B_0^{H^{\pm}L_2}$ which are related to the
mixing between charged Higgs with selectrons and smuons. The terms
proportional to $\Delta B_0^{H^{\pm}L_1}$ come from the diagrams of
Fig.~\ref{fig:ChargedScalarLoop-4}. The terms proportional to $\Delta
B_0^{H^{\pm}L_2}$ are easily obtained from these by replacing the
corresponding slepton lines and couplings. Notice that in
Fig.~\ref{fig:ChargedScalarLoop-4}~b) there is a contribution
proportional to $v_1/v$ that does not belong to this term. We
will show below that it will contribute to the 
$\Delta B_0^{G^{\pm}L_1}$ term.

\subsection{$\Delta B_0^{G^{\pm}L_1}$ and  $\Delta
B_0^{G^{\pm}L_2}$}

\begin{figure}[ht]
\begin{center}
\begin{tabular}{c}
\includegraphics[width=0.45\linewidth]{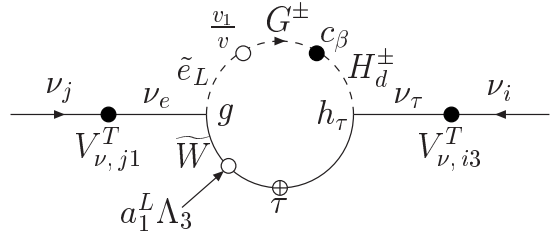}
\end{tabular}
\end{center}
\vspace{-3mm}
\caption{$G^{\pm}$ contribution to  $\Delta B_0^{G^{\pm}L_1}$}
\label{fig:ChargedScalarLoop-5}
\end{figure}

The graphs contributing to the $\Delta B_0^{G^{\pm}L_1}$ and $\Delta
B_0^{G^{\pm}L_2}$ terms are related to those of
Fig.~\ref{fig:ChargedScalarLoop-4}. They are given by
Fig.~\ref{fig:ChargedScalarLoop-5} and by the term proportional to
$v_1/v$ in Fig.~\ref{fig:ChargedScalarLoop-4} b), for the case of the
selectron.  The terms proportional to $\Delta B_0^{G^{\pm}L_2}$ are
easily obtained from these by replacing the corresponding slepton
lines and couplings.

\subsection{$\Delta B_0^{G^{\pm}\tilde\tau_1\tilde\tau_2}$}

\begin{figure}[ht]
\begin{center}
\begin{tabular}{cc}
\includegraphics[width=0.45\linewidth]{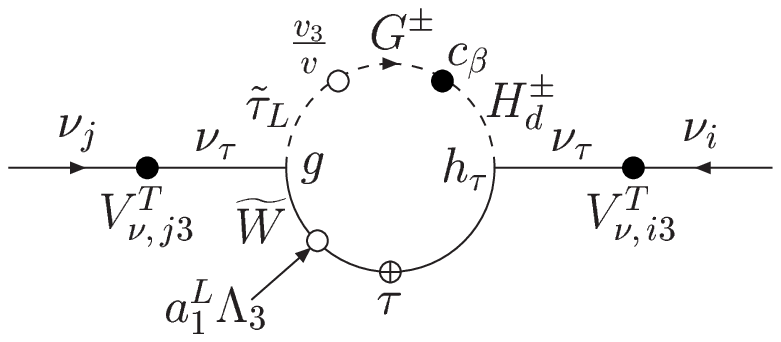}
&\includegraphics[width=0.45\linewidth]{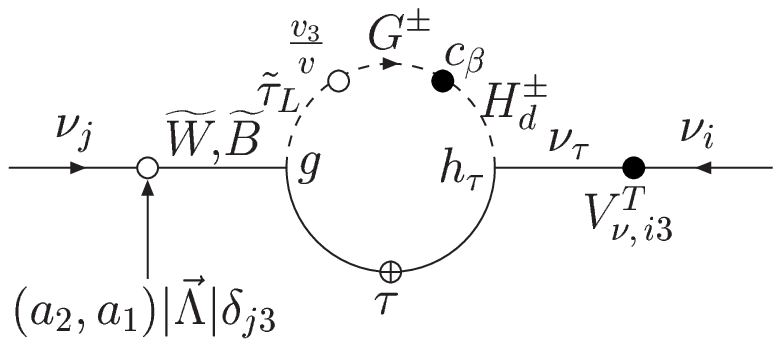}
\end{tabular}
\end{center}
\vspace{-3mm}
\caption{$G^{\pm}$ contribution to  $\Delta
  B_0^{G^{\pm}\tilde\tau_1\tilde\tau_2}$ }
\label{fig:ChargedScalarLoop-6}
\end{figure}

We now consider a more complicated term, the one proportional to
$\Delta B_0^{G^{\pm}\tilde\tau_1\tilde\tau_2}$. This term gives a
finite ultraviolet contribution and comes from the diagrams of
Fig.~\ref{fig:ChargedScalarLoop-6}, together with the parts of the
diagrams of Fig.~\ref{fig:ChargedScalarLoop-3} that are proportional
to $v_3/v$. Corresponding to the diagrams in
Fig.~\ref{fig:ChargedScalarLoop-6} proportional to $v_3/v$, there is
another set with $\tilde\tau_1$ and $\tilde\tau_2$ interchanged in the
usual way.

\subsection{$\Delta B_0^{G^{\pm}H^{\pm}\tilde\tau_1\tilde\tau_2}$}

\begin{figure}[ht]
\begin{center}
\begin{tabular}{cc}
\includegraphics[width=0.45\linewidth]{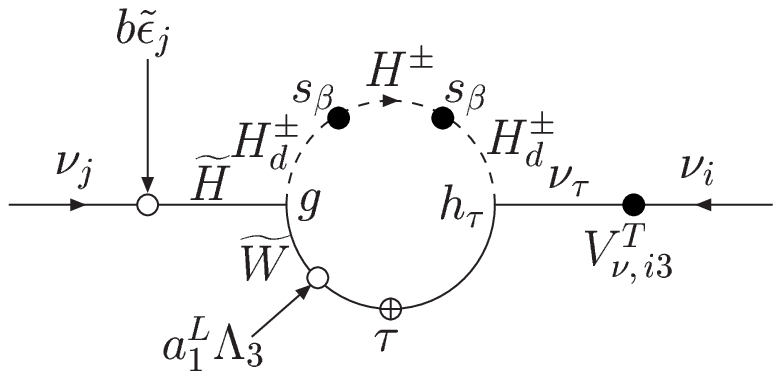}&
\includegraphics[width=0.45\linewidth]{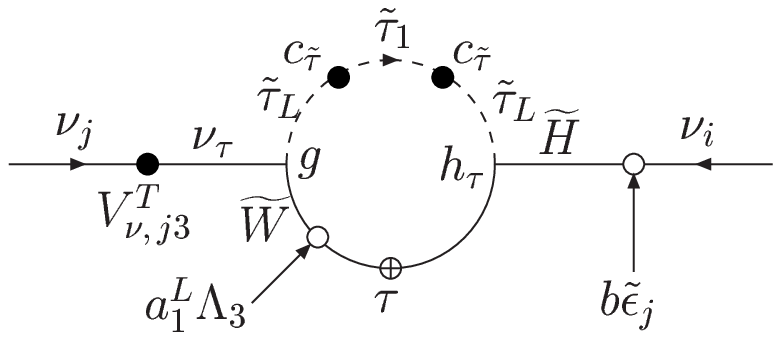}
\end{tabular}
\end{center}
\vspace{-3mm}
\caption{$H^{\pm}$ and $\tilde\tau_1$ contributions to  $\Delta
  B_0^{G^{\pm}H^{\pm}\tilde\tau_1\tilde\tau_2}$ }
\label{fig:ChargedScalarLoop-7}
\end{figure}

Let us consider finally the last term in Eq.~(\ref{ChargedHiggsLoop}),
the one proportional to $\Delta
B_0^{G^{\pm}H^{\pm}\tilde\tau_1\tilde\tau_2}$. This term gives
an ultraviolet finite contribution and comes from four diagrams. The
first two are those represented in
Fig.~\ref{fig:ChargedScalarLoop-7} corresponding to a $H^{\pm}$ and
$\tilde\tau_1$ propagating in the loop. The other two are obtained from
these with the replacements,
\begin{equation}
  \begin{tabular}{ccc}
    $H^{\pm}\rightarrow\ G^{\pm}$ &,& $\ s_{\beta}\rightarrow
    c_{\beta} $\\[+3mm] 
    $\tilde\tau_1\rightarrow\ \tilde\tau_2$ &,& $\
    s_{\tilde\tau}\rightarrow c_{\tilde\tau} $ 
  \end{tabular}
\end{equation}

\begin{thebibliography}{99}
  
\bibitem{Ahmad:2002jz}
Q.~R.~Ahmad {\it et al.}  [SNO Collaboration],
Phys.\ Rev.\ Lett.\ {\bf 89}, 011301 (2002) [arXiv:nucl-ex/0204008].

\bibitem{Maltoni:2002ni}
M.~Maltoni, T.~Schwetz, M.~A.~Tortola and J.~W.~Valle,
arXiv:hep-ph/0207227, to be published in Phys. Rev. D. This contains a 
full set of references on pre-KamLAND analyses of solar neutrino data. For
references to relevant experimental papers see \cite{Pakvasa:2003zv}

\bibitem{:2002dm}
K. Eguchi et al,  [KamLAND Collaboration],
arXiv:hep-ex/0212021.

\bibitem{Pakvasa:2003zv} 
For a recent review see S.~Pakvasa and
  J.~W.~Valle,
  arXiv:hep-ph/0301061. 

\bibitem{Maltoni:2002aw}
M.~Maltoni, T.~Schwetz and J.~W.~Valle,
arXiv:hep-ph/0212129, for a full set of references on post-KamLAND
analyses of solar neutrino data see ~\cite{Pakvasa:2003zv}.

\bibitem{Gonzalez-Garcia:1999aj}
M.~C.~Gonzalez-Garcia, P.~C.~de Holanda, C.~Pena-Garay and J.~W.~Valle,
Nucl.\ Phys.\ B {\bf 573}, 3 (2000) [arXiv:hep-ph/9906469].

\bibitem{Fukuda:1998mi}
Y.~Fukuda {\it et al.}  [Super-Kamiokande Collaboration],
Phys.\ Rev.\ Lett.\ {\bf 81}, 1562 (1998) [arXiv:hep-ex/9807003].

\bibitem{seesaw} M Gell-Mann, P Ramond, R. Slansky, in {\sl
    Supergravity}, ed. P.~van~Niewenhuizen and D.~Freedman (North
  Holland, 1979); T.~Yanagida, in {\sl KEK lectures}, ed.  O.~Sawada
  and A.~Sugamoto (KEK, 1979)


\bibitem{Mohapatra:1980yp}
R.~N.~Mohapatra and G.~Senjanovic,
Phys.\ Rev.\ D {\bf 23}, 165 (1981).

\bibitem{Schechter:1981cv}
J.~Schechter and J.~W.~Valle,
Phys.\ Rev.\ D {\bf 25}, 774 (1982).

\bibitem{Ross:1985yg}
G.~G.~Ross and J.~W.~Valle,
Phys.\ Lett.\ B {\bf 151} (1985) 375.
L.~J.~Hall and M.~Suzuki,
Nucl.\ Phys.\ B {\bf 231}, 419 (1984).
J.~R.~Ellis et al, 
Phys.\ Lett.\ B {\bf 150} (1985) 142.
C.~S.~Aulakh and R.~N.~Mohapatra,
Phys.\ Lett.\ B {\bf 119} (1982) 13
A.~Santamaria and J.~W.~Valle,
Phys.\ Lett.\ B {\bf 195}, 423 (1987).
Phys.\ Rev.\ Lett.\  {\bf 60}, 397 (1988).
Phys.\ Rev.\ D {\bf 39}, 1780 (1989).

\bibitem{Diaz:1997xc}
M.~A.~Diaz, J.~C.~Romao and J.~W.~Valle,
Nucl.\ Phys.\ B {\bf 524}, 23 (1998)
[arXiv:hep-ph/9706315].
For a review see J.~W.~Valle,
arXiv:hep-ph/9808292.

\bibitem{Bartl:2000yh}
A.~Bartl, W.~Porod, D.~Restrepo, J.~Romao and J.~W.~Valle,
Nucl.\ Phys.\ B {\bf 600} (2001) 39 [arXiv:hep-ph/0007157].

\bibitem{Porod:2000hv}
W.~Porod, M.~Hirsch, J.~Romao and J.~W.~Valle,
Phys.\ Rev.\ D {\bf 63}, 115004 (2001) [arXiv:hep-ph/0011248].

\bibitem{Mukhopadhyaya:1998xj}
B.~Mukhopadhyaya, S.~Roy and F.~Vissani,
Phys.\ Lett.\ B {\bf 443}, 191 (1998) [arXiv:hep-ph/9808265].

\bibitem{Choi:1999tq}
S.~Y.~Choi, E.~J.~Chun, S.~K.~Kang and J.~S.~Lee,
Phys.\ Rev.\ D {\bf 60}, 075002 (1999) [arXiv:hep-ph/9903465].

\bibitem{Hirsch:2002ys}
M.~Hirsch, W.~Porod, J.~C.~Romao and J.~W.~Valle,
Phys.\ Rev.\ D {\bf 66} (2002) 095006 [arXiv:hep-ph/0207334].

\bibitem{Restrepo:2001me}
D.~Restrepo, W.~Porod and J.~W.~Valle,
Phys.\ Rev.\ D {\bf 64}, 055011 (2001) [arXiv:hep-ph/0104040].



\bibitem{Allanach:1999bf}
B.~Allanach {\it et al.},
arXiv:hep-ph/9906224.



\bibitem{deCampos:1995av}
F.~de Campos et al,
Nucl.\ Phys.\ B {\bf 451}, 3 (1995) [arXiv:hep-ph/9502237].
Y.~Grossman and H.~E.~Haber,
Phys.\ Rev.\ D {\bf 63}, 075011 (2001) [arXiv:hep-ph/0005276].

\bibitem{Banks:1995by}
T.~Banks, Y.~Grossman, E.~Nardi and Y.~Nir,
Phys.\ Rev.\ D {\bf 52}, 5319 (1995) [arXiv:hep-ph/9505248].

\bibitem{deCarlos:1996du}
B.~de Carlos and P.~L.~White,
Phys.\ Rev.\ D {\bf 54}, 3427 (1996) [arXiv:hep-ph/9602381].

\bibitem{Akeroyd:1997iq}
A.~G.~Akeroyd et al,
Nucl.\ Phys.\ B {\bf 529}, 3 (1998) [arXiv:hep-ph/9707395].

\bibitem{BRpVrecent}
F. de Campos, O. J. P. \'Eboli, M. A. Garc\'{\i}a--Jare\~no and
J. W. F. Valle, {\sl Nucl. Phys. B} {\bf 546}, 33 (1999);
R. Kitano and K. Oda, {\sl Phys. Rev. D} {\bf 61}, 113001 (2000);
D. E. Kaplan and A. E. Nelson, {\sl JHEP} {\bf 0001}, 033 (2000);
C.--H. Chang and T.--F. Feng, {\sl Eur. Phys. J.} {\bf C12}, 137 (2000);
M. Frank, {\sl Phys. Rev. D} {\bf 62}, 015006 (2000);
F. Takayama and M. Yamaguchi, {\sl Phys. Lett. B} {\bf 476}, 116 (2000);
K. Choi, E. J. Chun and K. Hwang, {\sl Phys. Lett. B} {\bf 488}, 145
(2000);
M.A. Diaz, R.A. Lineros, M.A. Rivera, hep-ph/0210182;
F. De Campos, M.A. Diaz, O.J.P. Eboli, M.B. Magro, P.G. Mercadante,
{\sl Nucl. Phys. B} {\bf 623}, 47 (2002).

\bibitem{BRpVother}
A. S. Joshipura and M. Nowakowski, {\sl Phys. Rev. D} {\bf 51}, 2421
(1995);
G. Bhattacharyya, D. Choudhury and K. Sridhar, {\sl Phys. Lett. B}
{\bf 349}, 118 (1995);
A. Yu. Smirnov and F. Vissani, {\sl Nucl. Phys. B} {\bf 460}, 37 (1996);
J. C. Rom\~ao, F. de Campos, M. A. Garc\'{\i}a--Jare\~no, M. B. Magro
and J. W. F. Valle, {\sl Nucl. Phys. B} {\bf 482}, 3 (1996);
R. Hempfling, {\sl Nucl. Phys. B} {\bf 478}, 3 (1996);


\bibitem{BRpV_tau}
M. A. D\'{\i}az, J. Ferrandis, J. C. Rom\~ao and J. W. F. Valle,
{\sl Phys. Lett. B} {\bf 453}, 263 (1999);
A. G. Akeroyd, M. A. D\'{\i}az and J. W. F. Valle,
{\sl Phys. Lett. B} {\bf 441}, 224 (1998);
M. A. D\'{\i}az, E. Torrente--Lujan and J. W. F. Valle,
{\sl Nucl. Phys. B} {\bf 551}, 78 (1999); M. A. D\'{\i}az, J. Ferrandis,
J. C. Rom\~ao and J. W. F. Valle, {\sl Nucl. Phys. B} {\bf 590}, 3 (2000);
M. A. D\'{\i}az, D. A. Restrepo and J. W. F. Valle,
{\sl Nucl. Phys. B} {\bf 583}, 182 (2000);
M. A. D\'{\i}az, J. Ferrandis and J. W. F. Valle, {\sl Nucl. Phys. B}
{\bf 573}, 75 (2000).

\bibitem{BRpVmore}
S. Roy, B. Mukhopadhyaya, {\sl Phys. Rev. D} {\bf 55}, 7020 (1997);
K. Cheung, O.C.W. Kong, {\sl Phys. Rev. D} {\bf 64}, 095007 (2001);
T.F. Feng, X.Q. Li, {\sl Phys. Rev. D} {\bf 63}, 073006 (2001);
E.J. Chun, S.K. Kang, {\sl Phys. Rev. D} {\bf 61}, 075012 (2000);
J. Ferrandis, {\sl Phys. Rev. D} {\bf 60}, 095012 (1999);
A.G. Akeroyd, C. Liu, J. Song, {\sl Phys. Rev. D} {\bf 65}, 015008 (2002);
D. Suematsu, {\sl Phys. Lett. B} {\bf 506}, 131 (2001).



\bibitem{Hirsch:2000ef}
M.~Hirsch, M.~A.~Diaz, W.~Porod, J.~C.~Romao and J.~W.~Valle,
Phys.\ Rev.\ D {\bf 62}, 113008 (2000) [Erratum-ibid.\ D {\bf 65},
119901 (2002)] [arXiv:hep-ph/0004115];
J.~C.~Romao, M.~A.~Diaz, M.~Hirsch, W.~Porod and J.~W.~Valle,
Phys.\ Rev.\ D {\bf 61}, 071703 (2000) [arXiv:hep-ph/9907499].

\bibitem{Chun:2002vp}
E.~J.~Chun, D.~W.~Jung and J.~D.~Park,
arXiv:hep-ph/0211310.
F.~Borzumati and J.~S.~Lee,
Phys.\ Rev.\ D {\bf 66} (2002) 115012 [arXiv:hep-ph/0207184].
E.~J.~Chun and J.~S.~Lee,
Phys.\ Rev.\ D {\bf 60}, 075006 (1999) [arXiv:hep-ph/9811201].
A.~Abada, S.~Davidson and M.~Losada,
Phys.\ Rev.\ D {\bf 65} (2002) 075010 [arXiv:hep-ph/0111332].

\bibitem{Masiero:1990uj}
A.~Masiero and J.~W.~Valle,
Phys.\ Lett.\ B {\bf 251}, 273 (1990).

\bibitem{Romao:vu}
J.~C.~Romao, C.~A.~Santos and J.~W.~Valle,
Phys.\ Lett.\ B {\bf 288}, 311 (1992).

\bibitem{spo:2}
J.~C.~Romao and J.~W.~Valle,
Nucl.\ Phys.\ B {\bf 381} (1992) 87;
J.~C.~Romao, F.~de Campos and J.~W.~Valle,
Phys.\ Lett.\ B {\bf 292} (1992) 329 [arXiv:hep-ph/9207269];
M.~Shiraishi, I.~Umemura and K.~Yamamoto,
Phys.\ Lett.\ B {\bf 313} (1993) 89.
For bounds on spontaneous R-parity violation see P.~Abreu {\it et al.}
[DELPHI Collaboration],
Phys.\ Lett.\ B {\bf 502} (2001) 24 [arXiv:hep-ex/0102045]
and D.~Magalhaes Moraes,
CERN-THESIS-2002-021

\bibitem{Mira:2000gg}
J.~M.~Mira, E.~Nardi, D.~A.~Restrepo and J.~W.~Valle,
Phys.\ Lett.\ B {\bf 492} (2000) 81 [arXiv:hep-ph/0007266].

\bibitem{Giudice:1988yz}
G.~F.~Giudice and A.~Masiero,
Phys.\ Lett.\ B {\bf 206}, 480 (1988).

\bibitem{Nilles:1996ij}
H.~P.~Nilles and N.~Polonsky,
Nucl.\ Phys.\ B {\bf 484}, 33 (1997) [arXiv:hep-ph/9606388].

\bibitem{Hirsch:1998kc}
M.~Hirsch and J.~W.~Valle,
Nucl.\ Phys.\ B {\bf 557}, 60 (1999) [arXiv:hep-ph/9812463].

\bibitem{Now96}
M. Nowakowski and A. Pilaftsis, {\sl Nucl. Phys. B} {\bf 461}, 19
(1996).

\bibitem{Passarino:1978jh}
G.~Passarino and M.~J.~Veltman,
Nucl.\ Phys.\ B {\bf 160} (1979) 151

\bibitem{FFLP}Numerical algorithms for Passarino-Veltman functions 
have been published in: 
G.~J.~van Oldenborgh and J.~A.~Vermaseren,
Z.\ Phys.\ C {\bf 46} (1990) 425; 
T. Hahn, M. Perez-Victoria, Comput.Phys.Commun. 118 (1999) 153, see 
http://www.feynarts.de/looptools/;


\end{thebibliography}
\end{document}